\documentclass[%
 reprint,
amsmath,amssymb, aps,
]{revtex4-2}

\usepackage{graphicx}
\usepackage{dcolumn}
\usepackage{bm}

\begin{document}


\title{Experimental study of integrable turbulence in shallow water}
\author{Ivan Redor, Herv\'e Michallet, Nicolas Mordant, Eric Barth\'elemy}
\email{eric.barthelemy@grenoble-inp.fr}
\affiliation{Laboratoire des Ecoulements G\'eophysiques et Industriels, Universit\'e Grenoble Alpes, CNRS, Grenoble-INP,  F-38000 Grenoble, France}%

\date{\today}

\begin{abstract}
We analyze a set of bidirectional wave experiments in a linear wave flume of which some are conducive to integrable turbulence. In all experiments the wavemaker forcing is sinusoidal and the wave motion is recorded by seven high-resolution side-looking cameras. The periodic scattering transform is implemented and power spectral densities computed to discriminate linear wave motion states from integrable turbulence and soliton gas. Values of the wavemaker forcing Ursell number and relative amplitude are required to be above some threshold values for the integral turbulence to occur. 
Despite the unavoidable slow damping, soliton gases achieve stationary states because of the continuous energy input by the wavemaker. The statistical properties are given in terms of probability density distribution, skewness and kurtosis. The route to integrable turbulence, by the disorganization of the wavemaker induced sinusoidal wave motion, depends on the non-linearity of the waves but equally on the amplitude amplification and  reduction due to the wavemaker feedback on the wave field.
\end{abstract}

\maketitle
\section{Introduction}

Integrable turbulence is a fascinating topic of nonlinear physics. Integrable turbulence is theoretically and numerically described in the framework of integrable equations amongst which the KdV equation \cite{pelinovsky2006,carbone2016}, the Gardner equation \cite{Shurgalina2021}, and the 1D-nonlinear Schr\"odinger equation \cite{zakharov2009,Agafontsev2021}. In these systems an infinite number of degrees of freedom can be excited randomly.  As such there is no energy transfer between these modes and the word turbulence does not refer to the usual energy cascade between scales. Nevertheless, these systems can exhibit complex random behaviors that require a statistical description. Integrable turbulence applies to many fields of physics: hydrodynamics, optics, and plasmas. \cite{walczak2015optical,agafontsev2016integrable,soto2016integrable,randoux2016nonlinear,randoux2014intermittency,randoux2017optical,pelinovsky2006,dudley2019rogue,michel2020emergence,bonnefoy2020modulational,falcon2021experiments}.

The theory of integrable turbulence in water wave problems is found to be analytically tractable for two ``asymptotic'' situations. 
On the one hand when the waves are of small amplitude the expansion in powers of non linearity yields kinetic equations that model wave resonant interactions. In 2D situations such as for Kadomtsev-Petviashvili type equations \cite{zakharov2009} resonant interactions are three-wave resonances. It is known that in the case of the KdV equation the first nontrivial resonances are five-wave interactions but with zero amplitudes \cite{zakharov2014}. This result tends to indicate that integrable weak wave turbulence in 1D shallow water cases such as for the KdV equations is precluded.

On the other hand when the turbulence can be considered as a collection of solitons with random amplitudes and phases,
kinetic theories of rarefied soliton gas \citep{zakharov1971kinetic} or dense ones \citep{el2005kinetic,el2011kinetic,el2021soliton,congy2021soliton} can be derived . A soliton gas is thus a random state in which solitons behave as quasi-particles due to the fact that their collisions are elastic, only altering relative phases, and thus changing the mean phase speed \citep{zakharov1971kinetic}. Solitons in the shallow water framework are localized waves which propagate without change of shape due to a balance between linear dispersive effects that tend to flatten out any surface perturbations and nonlinear effects that steepen the fronts. Solitons are at the core of integrable dynamics of the KdV equation.

Empirical confirmation that soliton gases can be generated were given in optics \cite{schwache1997properties}, for deep water waves \cite{suret2020nonlinear} and for shallow water wave motion \cite{RedorPRL,RedorEIF}. In the experiments energy dissipation cannot be avoided and this seems at first glance strongly incompatible with the concept of integrability. However, \cite{RedorPRL} observed that due to a large scale separation between the nonlinear scale related to the (short) duration of soliton collisions and the (long) dissipative time scale, the dynamics is overall ruled by integrability. A stationary random soliton gas in a long wave flume in shallow water conditions can thus be sustained with continuous energy input by the wavemaker. Even though not labeled as soliton gases some 1D flume experiments of \cite{Ezersky2}, in which the wavemaker has a sinusoidal displacement, lead to random wave motions. Therefore, the role of the wavemaker in the outcome of these random stationnary wave states needs to be explored.

$50\,$ years have elapsed between the first theoretical description of soliton gases by \cite{zakharov1971kinetic} and the first hydrodynamic experiments. A possible reason is the requirement of highly resolved instruments to capture the space-time evolution of a random state. Another more fundamental issue relates to the difference between an infinite or periodic domain setting, usually used in theoretical approaches, and finite length experimental set-ups. This difference also combines with how initial conditions are easily prescribed in theory and numerics while boundary conditions are most of the time the only options at least in hydrodynamic experiments. A recent notable exception are the deep water soliton gas experiments by Suret {\it et al.} \cite{suret2020nonlinear} in which the Inverse Scattering Transform for the 1D Schr\"odinger equation is used to compute boundary conditions in a very long flume. In these experiments an ensemble of random spectral values associated to solitons are prescribed which then evolve towards a soliton gas.

The constraints on the experiments mentioned above, require to find other routes to the generation of integrable turbulence and soliton gases, an aspect investigated in the present work. In this context the question of the statistical properties of the stationary state of integrable turbulence needs to be addressed since it remains largely open and was mostly investigated by numerical studies~\cite{pelinovsky2006,carbone2016,Dutykh:2014dm,Pelinovsky:2017ce}. 

In the present study, we first present the experiments on random state wave motions and the data processing techniques (section \ref{ExpSet}). In section \ref{parametric}, we characterize the conditions for which such random states can be observed by using the Ursell number and the relative amplitude dimensionless number that define a phase diagram of our data. A statistical description of the random states is provided in section \ref{statistics}. Section \ref{transition} is dedicated to the description of 
the transition from wave motion to random states.

\section{Experimental setup}
\label{ExpSet}

The details of the experimental setup and data analysis tools can be found in \cite{Redorthesis} and some aspects are discussed in \cite{RedorPRL,RedorEIF} as well. We only recall here the main features of the experimental set-up.

\subsection{Wave flume and wave motion measurements} 
\begin{figure*}
\centering
\resizebox{\textwidth}{!}{\includegraphics[trim = 3cm 12.2cm 5.5cm 8.5cm, clip]{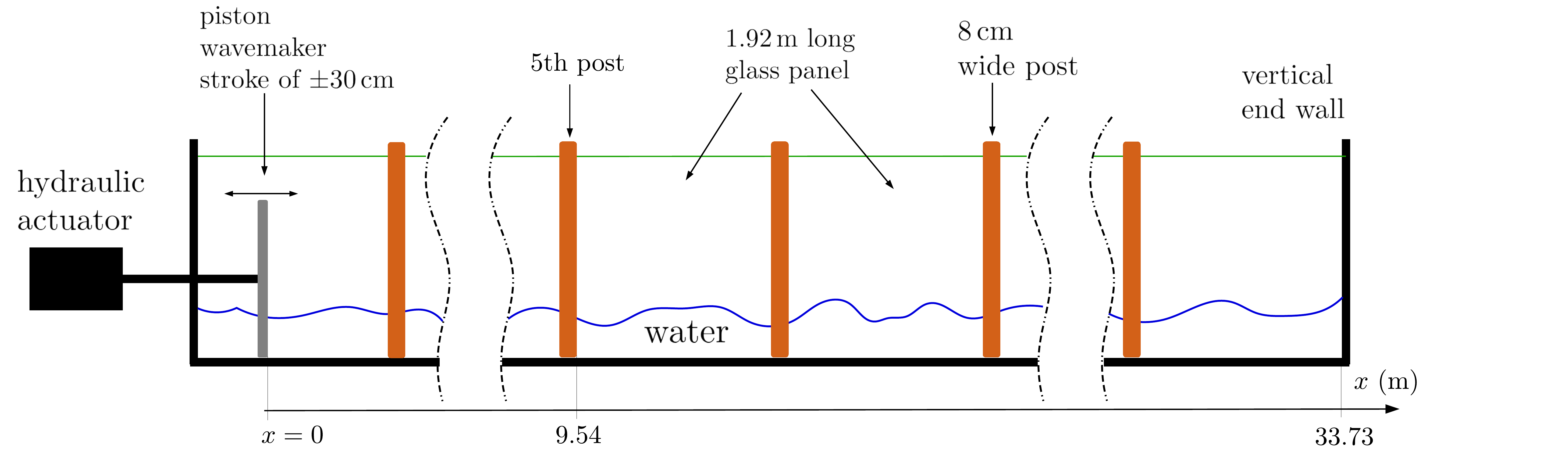}}
\caption{Schematics of the wave flume. The flume is $33.73$~m long, $1.3$~m deep and $55$~cm wide. 
Waves are generated by a piston-type wavemaker driven by a computer-controlled hydraulic actuator. The mean piston position is at $x=0$. 
$x=9.54\,$m is the right edge of the fifth post. The opposite end of the flume is a vertical wall on which waves reflect.
The side walls are made of glass with holding posts every $2$m.}
\label{flume}
\end{figure*}

\begin{figure}
\centering
\includegraphics[width=8cm]{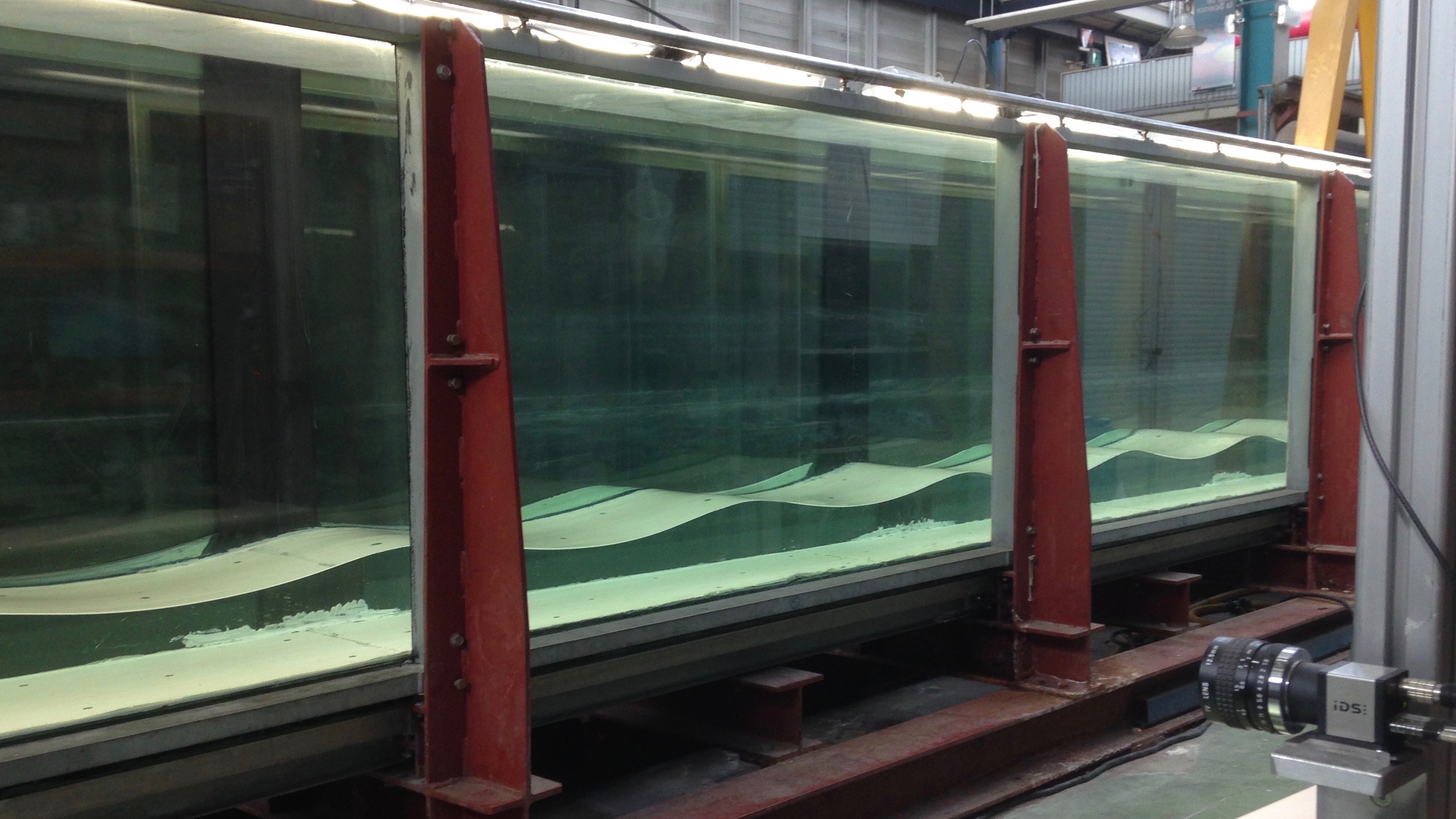}
\caption{Picture of the flume: the red-brown vertical posts and a USB camera (bottom right corner).}
\label{flume:pic}
\end{figure}

\begin{figure}
\centering
\resizebox{8cm}{!}{\includegraphics[trim = 0.2cm 1.5cm 3cm 0.7cm, clip]{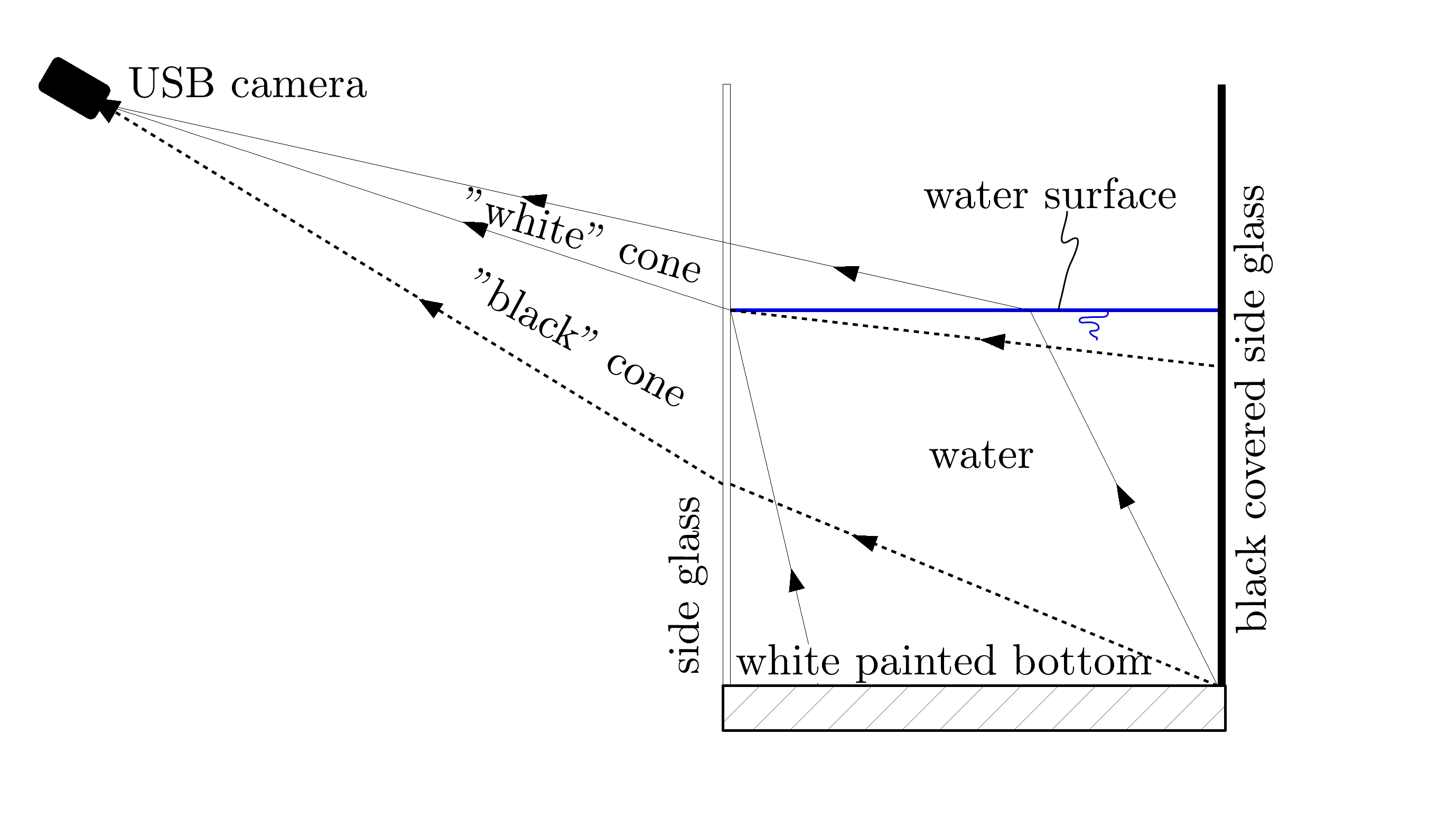}}
\caption{Schematics of the imaging of the surface elevation. 
The image refraction of the white bottom at the water surface creates a sharp contrast with the image refraction of the black back glass: 
the water contact line with the side glass is the boundary between the two images (see sample images in Fig.~\ref{measure2}). 
Dashed lines correspond to the light rays bounding the image of the black back glass panel as refracted by the vertical front water boundary with the glass. The plain lines correspond to the light rays bounding the image of the white bottom as refracted by the water surface.}
\label{measure}
\end{figure}

\begin{figure*}
\centering
\includegraphics[width=17cm]{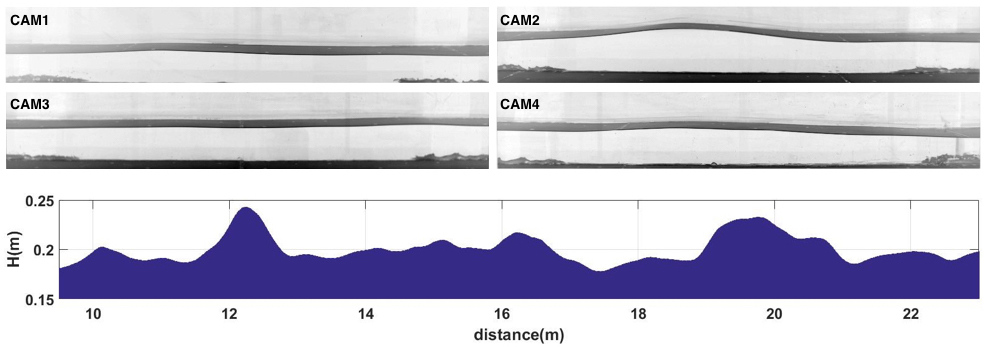}
\caption{Sample snapshots from the $4$ cameras the closest to the wavemaker (top). CAM1 for $9.54 \, {\rm m} \le x \le 11.46 \,  {\rm m}$; CAM2 for  $11.54 \, {\rm m} \le x \le 13.46 \,  {\rm m}$; CAM3 for  $13.54 \, {\rm m} \le x \le 15.46 \,  {\rm m}$; CAM4 for $15.54 \, {\rm m} \le x \le 17.46 \,  {\rm m}$. The images are in inverted gray scale levels and thus the water surface appears black. Bottom: reconstruction of the water surface elevation corresponding to the snapshots. The horizontal axis gives the distance to the wavemaker rest position. H is the measure of the fluid depth.}
\label{measure2}
\end{figure*}

Experiments are performed in the $33.73$~m-long and $55\,$cm-wide LEGI wave flume with side glass panels. 
A schematic of the flume is given in Fig.~\ref{flume} and a picture in Fig.~\ref{flume:pic}.
At one end the waves are generated by a piston-type wavemaker and opposite a vertical wall ends the flume, in a similar configuration to that of \cite{Ezersky,Ezersky2} but in a longer flume. 
Waves propagate back and forth in the flume reflecting on the wall and the wavemaker. They are damped by viscous dissipation in 
the boundary layers at the bottom and the sides of the flume. The side glass panels are $1.92$~m long and they are separated by $8$~cm wide posts that hold them. The water motion is video recorded through these side glass panels. The cameras have a full HD resolution ($1920\times 1080$ pixels) 
with pixel size corresponding to about $1$~mm in physical space. Each camera records the water contact line motion on the entire length of the side glass panel. Using $7$ synchronized cameras running at $20$~frames/s measurements of the water elevation over a $14$~m-long region located at the center of the flume are obtained. The raw images are rectified by using the image of a calibration grid that was placed in the flume against the front glass side panel. 
In order to improve the contrast of the images for a better accuracy of the measurements, 
the bottom of the flume is painted white and the back vertical panels painted black (see Fig.~\ref{measure}). 
By choosing adequately the angle of the camera, the image region just below the contact line appears black while the region above the contact line is white. This is due to the refraction of the light rays at the water surface as explained by the schematic in Fig.~\ref{measure}. 
This sharp contrast at the contact line allows us to achieve sub-pixel accuracy and a corresponding resolution better than $0.1$~mm (see \cite{Redorthesis, RedorEIF} for details of the water contact line measurement). Examples of snapshots recorded by the cameras are given in Fig.~\ref{measure2} together with the corresponding water free surface elevation. The set-up provides a time and space resolved measurement of the wave motion along roughly half the length of the wave flume. An example of a time-space reconstruction of the water elevation is shown in Fig.~\ref{fission1}.

\subsection{Wave generation} 

Our goal is to obtain integrable turbulence steady states such as soliton gases. Thus, one would like to generate as many solitons as possible with the wavemaker. In order to ultimately sustain a large number of solitons in the flume we take advantage i) of the well-known fission phenomenon in shallow water of a sine wave train into solitons as evidenced numerically by Zabusky \& Kruskal \citep{zabusky1965interaction} and observed experimentally by Zabusky \& Galvin \cite{Galvin1971} and in a more comprehensive way by Trillo {\it et. al} \cite{Trillo2016} and ii) of the amplitude amplification of the non-linear modes by their interactions with the moving wavemaker. 
Mention should be made of the experiments by \cite{Ezersky} who, by slightly detuning the wavemaker motion with respect to the periodic longitudinal seiching mode of the channel, were able to find a route towards the generation of integrable turbulence.

\begin{figure*}
\centering
\resizebox{0.8\textwidth}{!}{\includegraphics[trim = 2cm 0.8cm 1.8cm 2cm, clip]{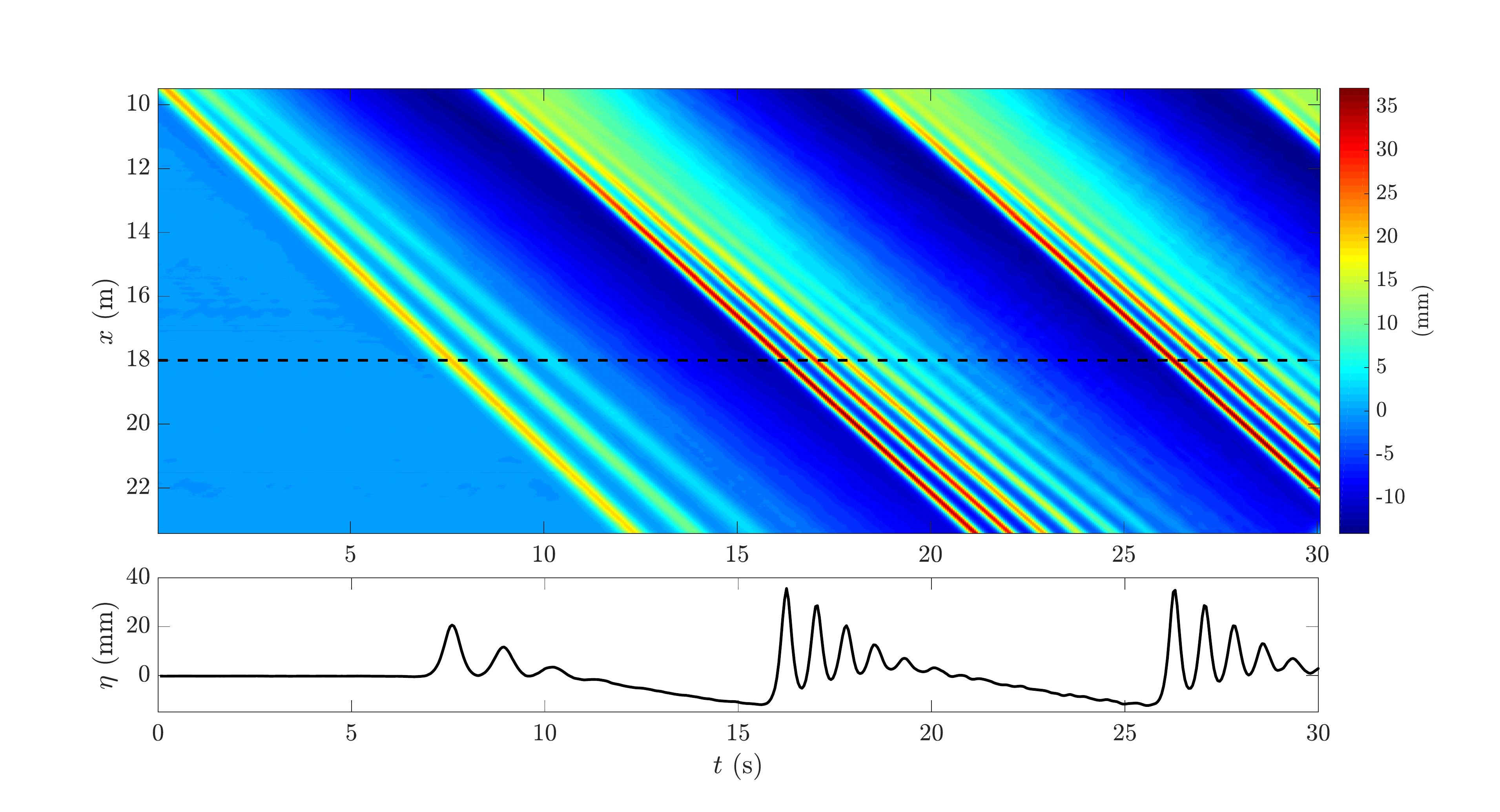}}
\caption{Fission of a sine wave into solitons. Water depth: $h=12$~cm; wavemaker sine motion period: $T=10$~s; wavemaker sine motion amplitude: $a=1.5$~cm. Upper panel: space-time representation of the wave field over the $14$~m field of view of the cameras and during the first $30$~s. The color-scale is given in mm with the zero corresponding to the free surface level at rest. Lower panel: time evolution at $x=18\,$m (dashed line in upper panel).}
\label{fission1}
\end{figure*}

Concerning point i), Fig.~\ref{fission1} is an experimental example of the fission of a sine wave train.
Indeed, a sine wave is not a stationary non-linear solution in shallow-water and integrability imposes an evolution towards a train of solitons and a non-solitonic background. 
The space-time representation in Fig.~\ref{fission1} is the full field recorded by the video cameras. The initial $30$~s of the record is presented corresponding to roughly $3$ wave periods. Leading is the first sine wave crest that evolves quite differently from the subsequent ones (bottom panel of Fig.~\ref{fission1}). The latter wave crests also undergo steepening due to nonlinear effects. At $x=10\,$m slight undulations are visible on theses wave crests corresponding to the regularization (or fission) of the forming shock by dispersive effects related. After a $23$~m propagation (right of the field of view) the soliton train is formed with $6$ clearly identified solitons of decreasing amplitudes. One can also note in top panel of Fig.~\ref{fission1} that the first soliton train which exits the field of view between $10$ and $17$~s, shows a different number of solitons compared to the subsequent wave trains. This is due to the fact that the first soliton train propagates on the rest/initial water level, while the subsequent trains seem to propagate on a lower (negative) water level since part of initial fluid volume is removed to generate the  train of solitons.

Concerning point ii) the amplification of the non-linear modes by the energy input provided to waves incoming at the wavemaker will be discussed in detail in section~\ref{transition}.

\subsection{Spectral and periodic scattering data analysis} 
\subsubsection{Fourier power spectrum}
\begin{figure}
\centering
\resizebox{0.45\textwidth}{!}{\includegraphics[trim = 0cm 0cm 0cm 0cm, clip]{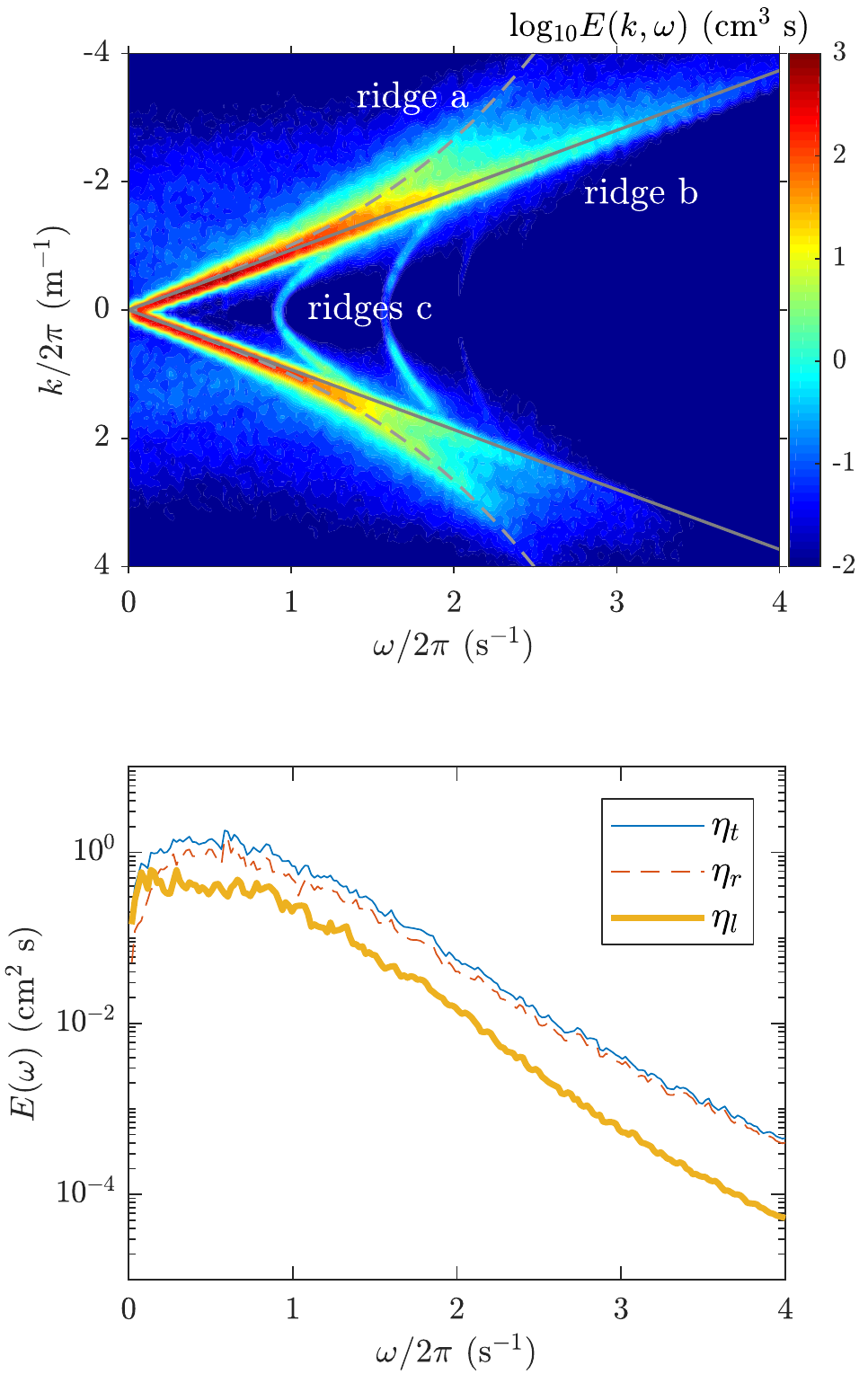}}
\caption{Power density spectrum of a soliton gas. Water depth $h=12$~cm; forcing frequency $f=0.6$~Hz; forcing sine amplitude $a=1.5$~cm; Ursell number $U=0.53$. Top panel: 2D Fourier transform. Bottom panel: 1D frequency Fourier transform; blue plain line: total signal ($\eta_t$); red dashed line: right-running waves ($\eta_r$); yellow bold line: left-running waves ($\eta_l$).}
\label{spectre}
\end{figure}

A standard tool to study wave propagation is the 2D Fourier power spectral density \citep{papoulis} defined here as:
\begin{equation}
E(k,\omega)=\frac{1}{2\pi L \, T}\, \Bigg \langle \, \left|\iint \eta(x,t)\, e^{i(kx+\omega t)} \,  {\rm d}t \, {\rm d}x \right|^2  \Bigg \rangle
\end{equation}
where $\eta(x,t)$ is the free surface displacement. The space integral for $x$ covers the field of view of length $L$ of the cameras and the time $t$ integral spans a time window of duration $T$. It is notable that the Fourier transforms are actually discrete Fourier transforms due to discrete sampling in space and time. The statistical average represented by $\langle . \rangle$, is an average over successive temporal windows using the standard Welch method. An example of such a spectrum is shown in Fig.~\ref{spectre}(a). Of note several ridges of energy higher than that of the background with the following interpretations: 
\begin{itemize}
\item ridge a: signature of weak dispersive shallow water waves following the Airy dispersion relation
\begin{equation}
\omega^2 = g \, k \, \tanh(k \, h).
\end{equation}
These dispersive waves originate from the bound waves of the wavemaker monochromatic sinusoidal wave forcing and from the weak non integrable effects during soliton collisions \cite{Redorthesis,RedorPRL};
\item ridge b: signature of shallow water non-linear waves which Fourier modes all travel at a velocity close to $c_0 = \sqrt{g \, h}$.
\item ridges c: signature of  transverse waves. The uni-nodal transverse wave would be at a frequency of $0.9\,$Hz. 
The energy of the uni-nodal transverse waves is three orders of magnitude smaller than that of longitudinal waves.
\end{itemize}

\begin{figure}
\centering
\resizebox{0.45\textwidth}{!}{\includegraphics[trim = 0cm 1.1cm 1.3cm 0.5cm, clip]{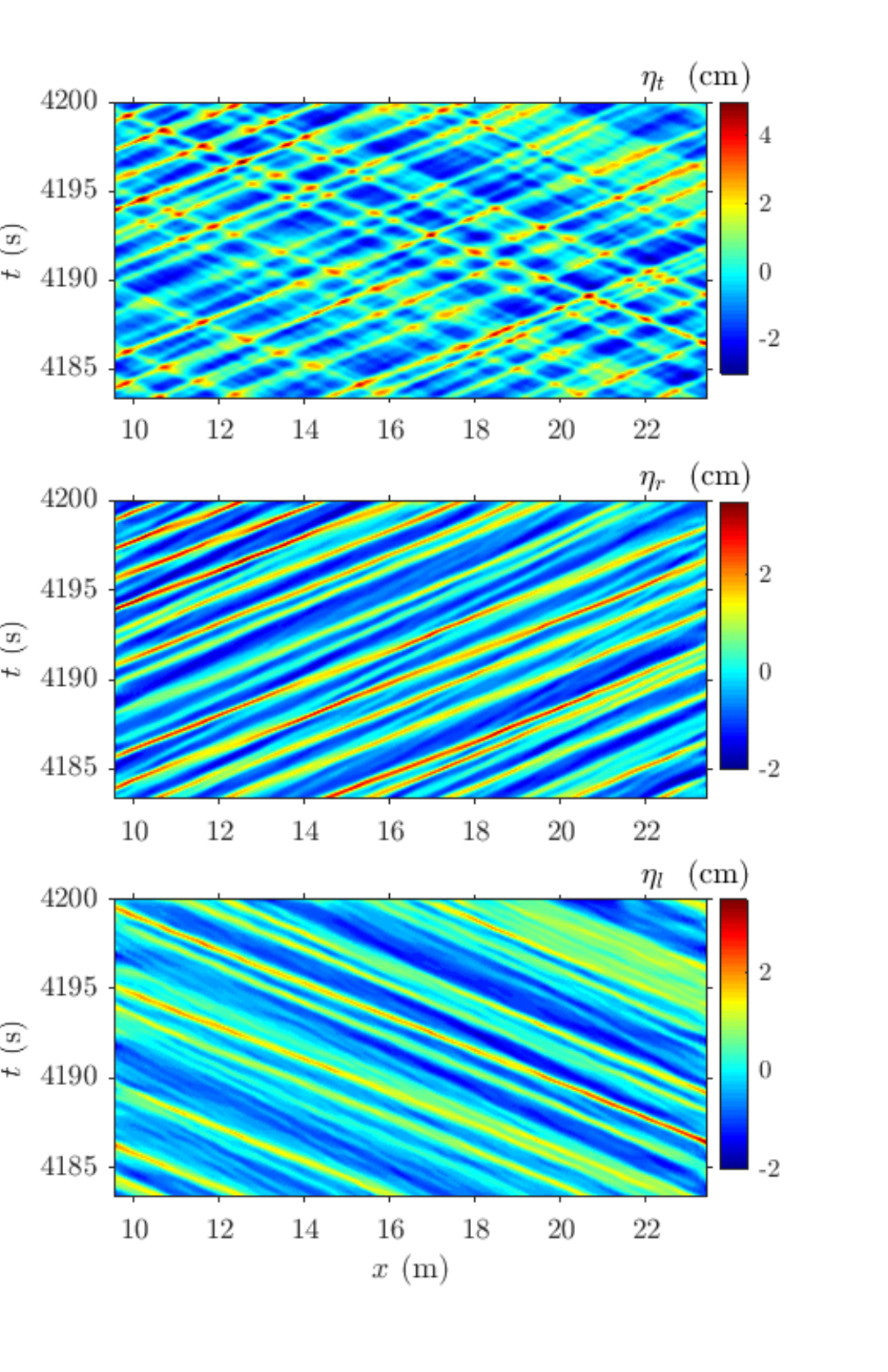}}
\caption{Splitting of the bidirectional wave field into unidirectional fields, case of a soliton gas. Top panel: the total wave motion field $\eta_t$; middle panel: the right-running waves $\eta_r$; bottom panel: the left-running waves $\eta_l$. Water depth $h=12$~cm;   wavemaker sine motion period: $T=10$~s; wavemaker sine motion amplitude: $a=1.5$~cm.
Color scale in centimeters.}
\label{separation}
\end{figure}

Waves corresponding to the upper-right quadrant of the Fourier space ($k<0$ and $\omega>0$) propagate to the right and those corresponding to the lower-right quadrant ($k>0$ and $\omega>0$) propagate to the left. Thus, by selecting a specific quadrant $(k,\omega)$ of the Fourier transform and inverse transforming back to the physical space, it is possible to separate waves going into opposite directions (see an example in Fig.~\ref{separation}). An alternative for this separation is the Radon transform as discussed in \cite{RedorEIF}. We also use the time 1D frequency Fourier power spectrum density for the analysis of the signals. An example of such a spectrum is shown in Fig.~\ref{spectre}(b) either for the full signal, for the split into right- and left-propagating waves.
The case described in Fig.~\ref{spectre} and \ref{separation} and analyzed in \cite{RedorPRL} is typically an example of integrable turbulence containing a significant number of solitons. Solitons are responsible for ``ridge b'' of Fig.~\ref{spectre} and they also clearly leave straight line signatures in the $(x,t)$ plane of Fig.~\ref{separation}. This was discussed in detail in  \cite{RedorEIF}.

\subsubsection{Periodic Scattering Transform}
While the Fourier Transform (FT) is the adequate processing tool to study the propagation of linear waves, it is unfit to discriminate an ensemble of nonlinear coherent waves such as solitons. Because of the nonlinear interactions, the FT spectral components of evolving nonlinear waves are not invariants as for linear waves.

Gardner~{\it et al.} \citep{Gardner67} made a significant leap forward with the Direct Scattering Transform (DST) of the KdV equation that extracts the spectrum of the associated Schr\"odinger equation which potential is the nonlinear wave signal to be analyzed. This method decomposes a time series into nonlinear solitonic modes that evolve independently in time and space without change of shape along with radiating shallow water weakly nonlinear waves. Once the time independent spectrum of nonlinear modes is determined, the signal can be reconstructed at any time by the Inverse Scattering Transform (IST). 
DST theory for the case of a localized initial condition  \citep{Gardner67} differs from the much more complex case of the spatially periodic initial condition known as the finite band theory \citep{novikov1974periodic,ablowitz1981solitons,novikov1984theory,christov2012hidden}. 
In our experimental case we are more concerned with the periodic case due to our configuration in which the waves propagate back and forth in the flume. Although the waves in the present experiments are not in most cases periodic, the wave motion is however confined in space and does not decay at infinity. This approach has been used by \cite{Costa} to analyze oceanic field data.

We implemented the Periodic Scattering Transform algorithm (PST) for the KdV equation developed by Osborne \cite{Osborne,Osborne94,Osborne86} in a program to compute the nonlinear spectrum of the experimental free surface records. In the periodic case, the non linear modes are described by hyper-elliptic functions. These waves similar to cnoidal waves  are characterized by the so-called spectral modulus $m$ which quantifies the level of nonlinearity and which is an output of the PST \cite{Osborne94}.  For a vanishing small modulus the modes are close to sine waves. For $m$ very close to $1$, the modes appear as localized pulses in the periodic box that resemble solitons. The delineation between solitonic modes and radiation modes thus breaks down to the choice of the threshold modulus. This threshold modulus also defines the reference depth $h_{ref}$ \cite{Osborne,Osborne94,Osborne86} on which these solitonic modes propagate. Osborne \cite{Osborne} suggests that modes with $m>0.99$ can be considered as solitons a definition also used by \cite{Trillo2016}. We will not discuss here the details of our implementation of the PST which is described in \cite{Redorthesis} with different validation cases and an analysis of various limitations since this method has been validated various times \citep{christov2012hidden,Trillo2016}. The main output of the PST is the spectrum that lists the nonlinear modes and their moduli. 

It must be noted that the present implementation of the PST is related to the KdV equation framework which describes unidirectional wave propagation only. Thus, for a consistent application of the PST to our measurements, we first separate left and right-propagating waves using the $(k,\omega)$ Fourier spectrum as described above and then apply the PST either to left or right-propagating waves. 

\begin{figure*}
\centering
\resizebox{0.65\textwidth}{!}{\includegraphics[trim = 1.9cm 4.8cm 2cm 4.8cm, clip]{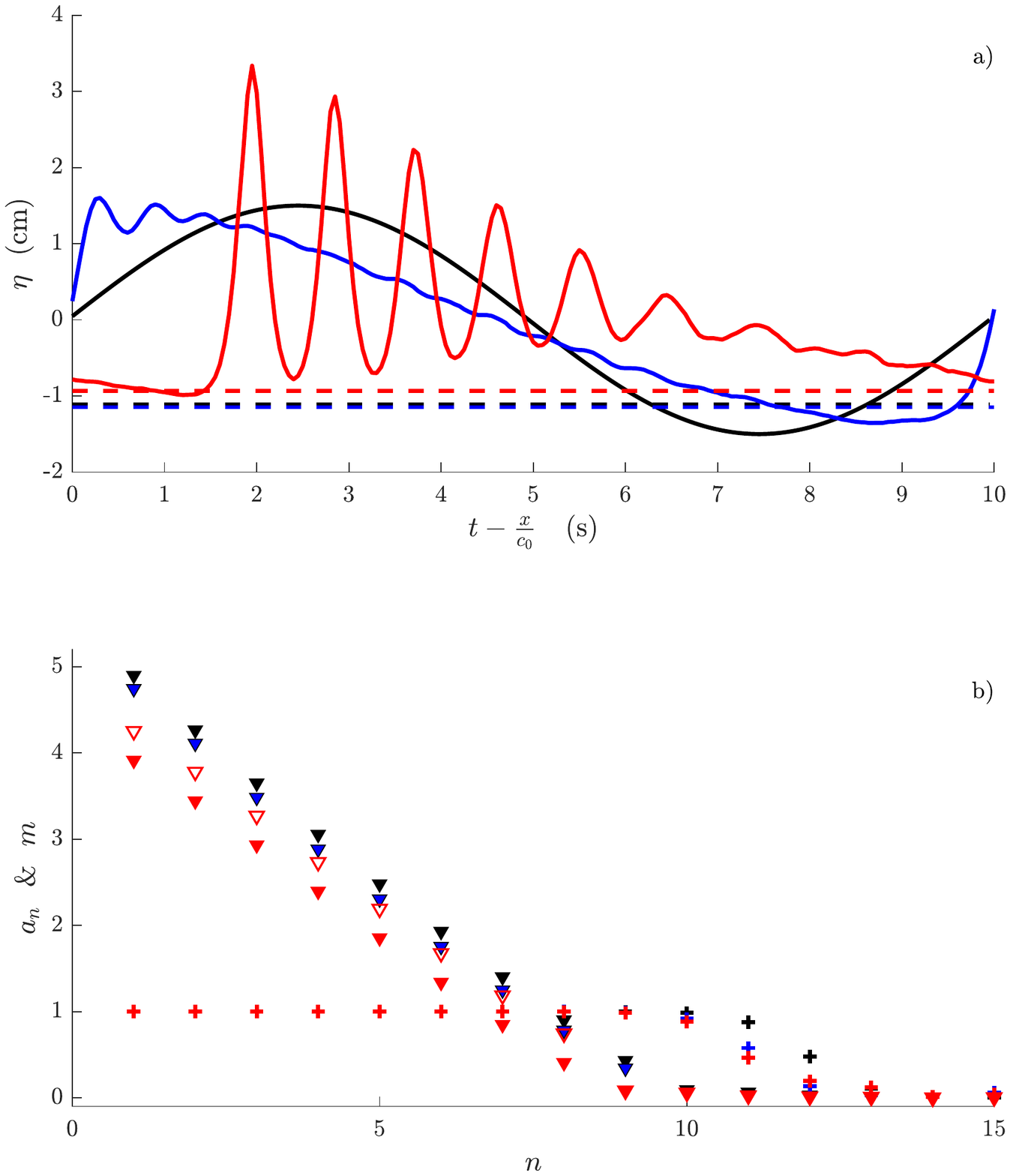}}
\caption{PST analysis of the fission of a sine wave ($U=19.4$) into solitons. Water depth $h=12$~cm. Sine motion forcing period $T=10$~s and amplitude $a=1.5$~cm.
Top panel a): free surface elevation $\eta$ at different positions along the flume, black: at the wavemaker $x=0$; blue: $x=10$~m; red: $x=23$~m; The dashed lines are the reference levels at each $x$ locations as computed by the PST with the choice of $m=0.99$ threshold modulus. 
Bottom panel b):  PST modal amplitudes $a_n$ ($\bigtriangledown$) and module $m$ ($+$) of the different modes of number $n$ at different position along the flume (same color coding as top panel). Full symbols: threshold modulus at $m=0.99$; empty symbols: threshold modulus at $m=0.985$.}
\label{exPST} 
\end{figure*}

Among the issues related to PST and discussed in~\cite{Redorthesis} one can mention the case of a signal containing solitonic modes of equal or very close amplitudes for which the detection of the modes is challenging since the eigenvalues of the spectrum are very close. Another issue raised in~\cite{Redorthesis} is the impact of the duration of the signal sample on the number of solitons and the reference depth especially for a soliton gas. Indeed, the longer the signal the more solitonic modes will be detected and therefore the lower the reference level. A last issue is related to the arbitrary threshold and the impact of solitons which moduli are very close to the threshold modulus. This issue is illustrated in Fig.~\ref{exPST} related to the fission of the sine into solitons as already discussed above. 
The PST has been applied to the three measured wave elevation profiles shown in Fig.~\ref{exPST}(a). The corresponding moduli and wave amplitudes (triangles with same colors) as computed with the PST are plotted in Fig.~\ref{exPST}(b). A threshold modulus of $0.99$ yields $9$ solitonic modes at $x=0$ and $x=10$~m but only $8$ solitons for the signal at $x=23$~m. Indeed, the modulus of the $9$th mode at $x=23$~m drops slightly below the threshold modulus $0.99$. Would the threshold be lowered down to $0.985$, the $9$th mode at $x=23$~m would be considered as a soliton. The soliton amplitudes are seen to be quite close in the three cases with a small decay with distance due to dissipation. Moreover this reduction of amplitude with $x$ is enhanced by the fact that since the $9$th mode is no longer a soliton at $x=23$~m the reference level automatically raises reducing the soliton amplitudes. 
The amplitudes with threshold at $0.985$, at $x=23$~m, are shown in Fig.~\ref{exPST}(b) as empty triangles and are seen to be larger than the ones for the $0.99$ threshold (filled triangles) and significantly closer to the ones obtained at $0$~m and $10$~m except for a slight decay due to dissipation. Although both sets of amplitude are globally consistent with the measurement, the empty symbols are closer to the measurements, at least for the large solitons. This illustrates the effect of the choice of a threshold modulus in the case of a large number of nonlinear modes which amplitudes and moduli have a wide distribution.
%
%
\section{Parametric study of the stationary regime}
\label{parametric}
\begin{figure*}
\centering \resizebox{0.75\textwidth}{!}{\includegraphics[trim = 2.5cm 4.3cm 2.7cm 4.3cm, clip]{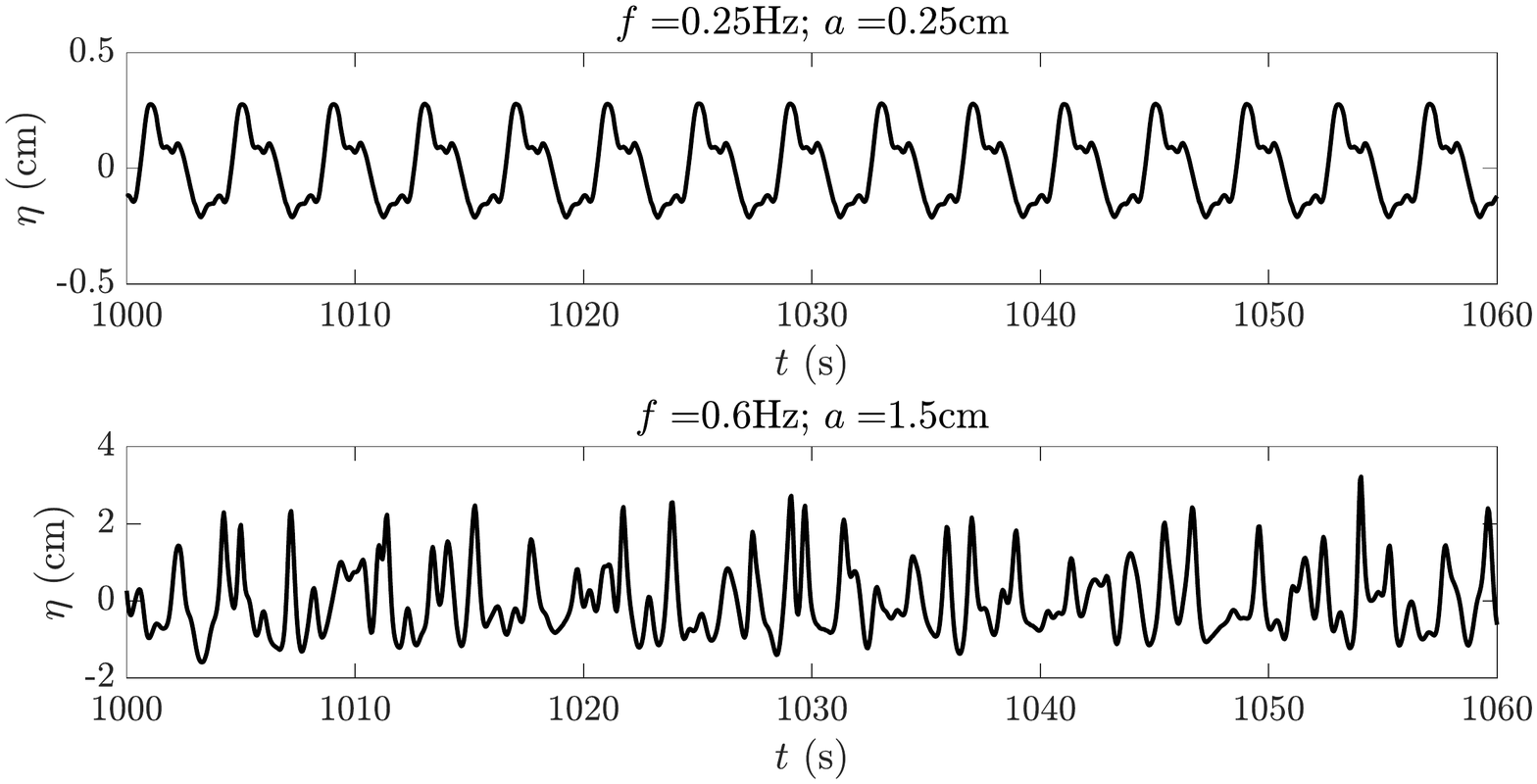}}
\caption{Time series of the free surface displacement forced by a sine wave prescribed at the wavemaker. Water depth is $h = 12 \,$cm; top panel: wavemaker frequency $f=0.25\,$Hz, amplitude $a=0.25\;$cm, Ursell number $U= 0.52$; bottom panel: wavemaker frequency $f=0.6\,$Hz, amplitude $a=1.5\;$cm, Ursell number $U= 0.53$ (these cases are also  in Fig.~\ref{PSDU}).}
\label{regimes}
\end{figure*}

In the previous sections we discussed the transient evolution of a sine wave and recalled that some conditions produce an integrable turbulence. However, is it the fate of all wave conditions to evolve into a random wave motion state? Two examples of wave motion recordings are plotted in Fig.~\ref{regimes} corresponding to two different regimes. These recordings are taken well after the start of the experiment, at roughly $16\,$mn. As time goes, one remains periodic while the second becomes disorganized or random. What makes these $2$ cases different?

The sine wave forcing at the wavemaker depends on $3$ physical parameters: the amplitude $a$ of the sinusoidal wave, the period $T$ (or the wavenumber $k = 2\pi/(c_0 \, T)$ and the water depth $h$. The space of parameters was explored by changing the values of these $3$ parameters. 
The amplitude was typically varied in a range $a\in[0.125,1.5]$~cm, the period in a range $T\in[1.6,5.5]$~s (or the frequency $f\in[0.18,0.6]$~Hz) and 
the water depth was set to the values $10$, $12$, $16$ and $20$~cm. 

One of the relevant non-dimensional number in the shallow water context is the Ursell number \cite{ursell1953long}.
It is even the only dimensionless number for KdV unidirectional wave motion dynamics.
The Ursell number is the ratio in order of magnitude of the nonlinear to dispersive terms of the KdV equation. The dimensional version of the KdV equation is,
\begin{equation}
\partial_t \eta + c_0 \, \partial_x \eta + \frac{3 \, c_0}{2 \, h} \, \eta \, \partial_x \eta + \frac{h^2 \, c_0}{6} \, \partial_{xxx} \eta = 0. \label{kdv:dim}
\end{equation}
Consider the following scaled dependent and independent variables \citep{Dutykh:2014dm},
\begin{equation}
\eta \leftarrow   \frac{3}{2} \,  \frac{h^2}{l^2 a} \, \eta \ \ \ \ \ \  x \leftarrow \frac{\sqrt{6}}{h}\,(x-c_0 \, t)  \ \ \ \ \ \  t \leftarrow \frac{\sqrt{6}}{h} \, c_0 \, t
\end{equation}
where $h$ is the water depth at rest, $a$ a characteristic vertical length scale (the amplitude of the forcing wave for instance) 
and $l$ a horizontal length scale (the forcing wave wave length).
Then equation (\ref{kdv:dim}) scales to,
\begin{eqnarray}
& \partial_t \eta + R \, \eta \, \partial_x \eta +  \partial_{xxx} \eta = 0 \label{kdv:adim} \\
& R = \frac{a \, l^2}{h^3}
\end{eqnarray}
where $R$ is a Ursell number.
Different pre-factors can be appended to $R$. Hereafter we use the following definition of the Ursell number used by \cite{osborne1994laboratory},
\begin{equation}
U=\frac{3}{4} \, \frac{a}{k^2 \, h^3} \label{def_ursell}
\end{equation}
where $k = 2 \, \pi /l$ is the wave number. This writing is exactly the ratio of the amplitudes of the second-order to the first-order term in the Stokes expansion of the water wave problem. For a time series the Ursell number would be,
\begin{equation}
U = \frac{3}{16\pi^2} \, \frac{a \, c_0^2 \, T^2}{h^3}  \label{def_ursellT}
\end{equation}
Equation (\ref{kdv:adim}) shows that for very small $U$ the equation becomes linear and dispersive, the so-called Airy equation. 
Under this condition the forcing wave remains linear but disperses (different frequency components propagate at different speeds) as it propagates and no soliton emerges. For large values of $U$ equation (\ref{kdv:adim}) becomes non-linear of the Burgers type. A sinusoidal forcing wave that fulfills such condition will undergo nonlinear steepening up to the gradient catastrophe producing a steep front face (shock wave). At that point the wave front face characteristic length is small and dispersion comes into play. Dispersion forces the shock wave to fission into a train of solitons \citep{zabusky1965interaction,Kurkina:2016fb,Trillo2016}. The Ursell number of the forcing wave therefore indicates how many solitons are expected \cite{Trillo2016} . Since our set-up does not allow for initial condition recurrence to take place because of the end wall reflection, the Ursell number also measures how disorganized the regime is. Indeed, solitons will reflect back on the end wall and the wavemaker, interacting with others, generating phase shifts and therefore possibly disorganizing the initial periodicity. 
\begin{figure*}
\centering
 \resizebox{\textwidth}{!}{\includegraphics[trim = 2.7cm 2.0cm 2.7cm 1cm, clip]{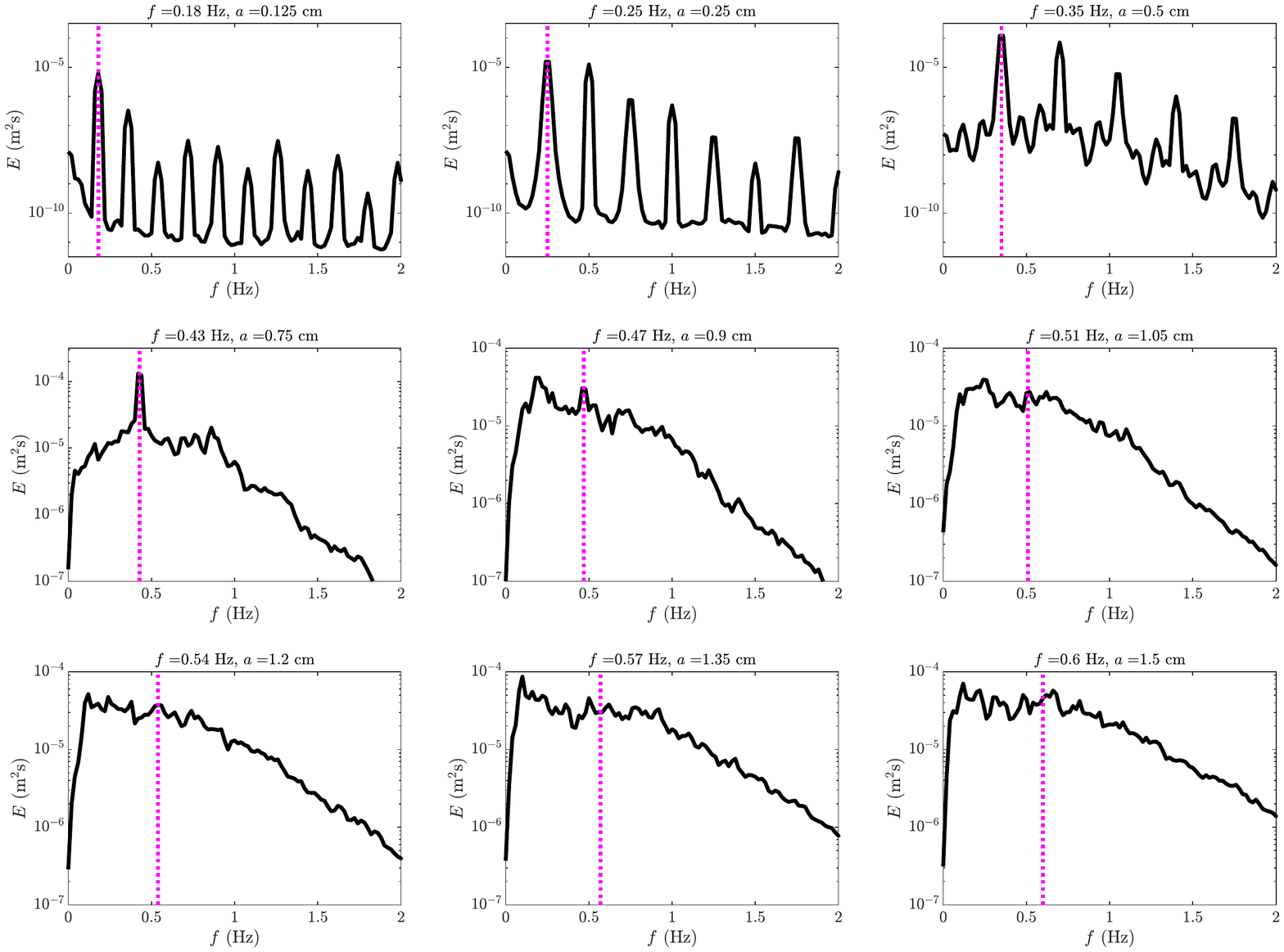}}
\caption{Power density spectra of the surface elevation for experiments with the same Ursell number, $U\simeq 0.53$, and differing values of the forcing amplitude $a$ and frequency $f$. Left-running waves at $x=40\,$m. Depth, $h = 12\,$cm. Vertical dashed line indicates the wavemaker forcing frequency. From top to bottom and left to right: $I_m = 2.54$, $I_m = 3.93$, $I_m = 3.38$, $I_m = 1.12$, $I_m = 0.40$, $I_m = 0.19$, $I_m = 0.16$, $I_m = 0.20$, $I_m = 0.25$ (see (\ref{index:random}) in text for the definition of $I_m$).}
\label{PSDU}
\end{figure*}

\begin{figure*}
\centering
\resizebox{0.9\textwidth}{!}{\includegraphics[trim = 2cm 8.5cm 2cm 8.2cm, clip]{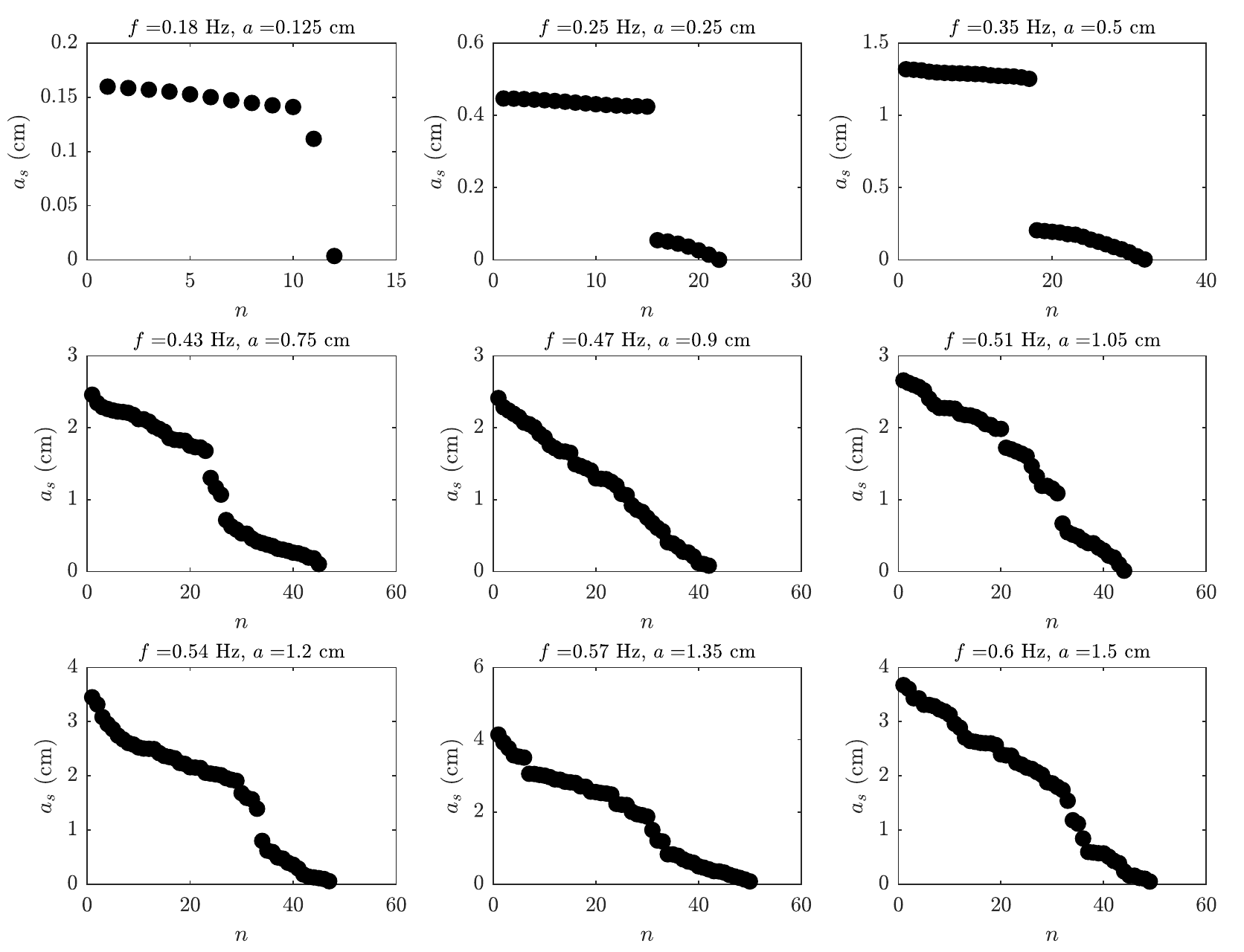}}
\caption{Solitonic mode amplitude $a_s$ distribution versus mode number $n$ computed by PST; same data as in figure~\ref{PSDU}: $U\simeq 0.53$, left-running waves at $x=40\,$m, $h = 12\,$cm.}
\label{PISTU}
\end{figure*}

However, we observed that the sole value of the forcing wave Ursell number $U$ is insufficient to discriminate between the periodic and random states. Fig.~\ref{PSDU} shows the frequency spectra of $9$ experiments for the left-running component of the wave motion at $x=40\,$m which is the best compromise between soliton separation after fission and dissipation. All these experiments have very close Ursell number values of $U = 0.53$, but distinct values of the frequency $f$ and amplitude $a$. Signatures of different stationary states are observed from very periodic ones at low forcing frequency and low forcing amplitude to random ones at high forcing frequency and high forcing amplitude and going through continuously varying power spectra shape in between. This observation shows that another dimensionless parameter must be taken into account to sort the different states out. Fig.~\ref{PISTU} shows the corresponding distribution of non-linear mode amplitudes given by the PST. The periodic cases (3 top subplots of Fig.~\ref {PISTU} and correspondingly in Fig.~\ref{PSDU}) exhibit sets of non-linear modes of nearly constant amplitudes indicating that these cases remain strongly organized. For instance the case $f=0.18\,$Hz for a $60$~s wave motion recording corresponds to roughly $10$ periods of the forcing wave for which the PST gives $10$ soliton modes of $0.15\,$cm amplitudes and hardly any other modes. A small soliton of $0.15\,$cm amplitude is locked to each wave period. By contrast the soliton modes in the random cases are much more numerous and their amplitudes distributed over a large range indicating indeed that these cases are random. These states (3 bottom subplots of Fig.~\ref{PISTU} and correspondingly in Figure~\ref{PSDU}) are considered to be what is called integrable turbulence \citep{RedorPRL}. Discussion of the amplitude distribution is post-poned to section~\ref{statistics}.

\begin{figure*}
\centering
\resizebox{0.9\textwidth}{!}{\includegraphics[trim = 2.8cm 6.3cm 3.3cm 6.5cm, clip]{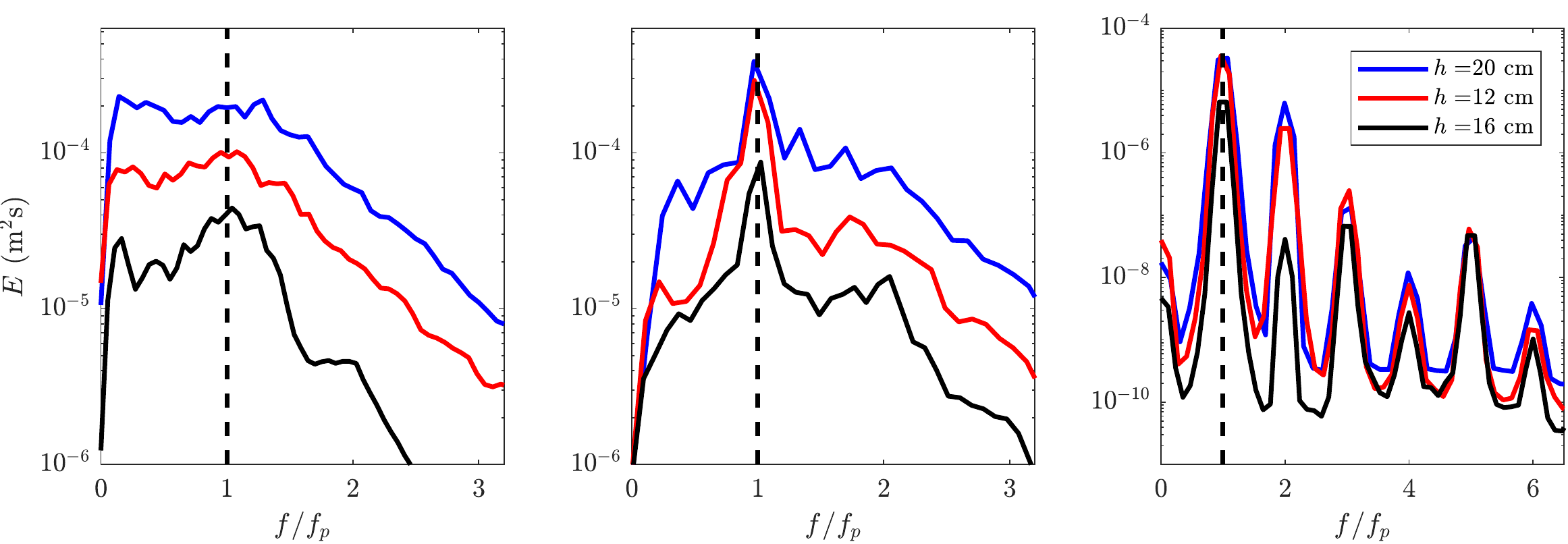}}
\caption{Examples of PSD of experiments with the same Ursell number $U$ and same values of $a/h$ and different water depth $h$.
left panel: $U=0.31$ \& $a/h=0.105$; middle panel: $U=0.52$ \& $a/h=0.0625$; right panel: $U=0.28$ \& $a/h=0.021$.}
\label{UAH}
\end{figure*}

To delineate integrable turbulence from other stationary states, the forcing wave non-linearity ratio $a/h$ is necessary where $a$ is the wavemaker amplitude.
Fig.~\ref{UAH} shows that states with identical $U$ and $a/h$ values but with differing $h$ have
power spectra of similar shape either continuous, peaked or mixed. 
It therefore appears that the forcing wave dimensionless numbers $U$ and $a/h$ are the main dimensionless numbers to discriminate between various stationary wave motion states.

\begin{figure*}
\centering
\resizebox{0.85\textwidth}{!}{\includegraphics[trim = 2.2cm 0.8cm 2.1cm 1.2cm, clip]{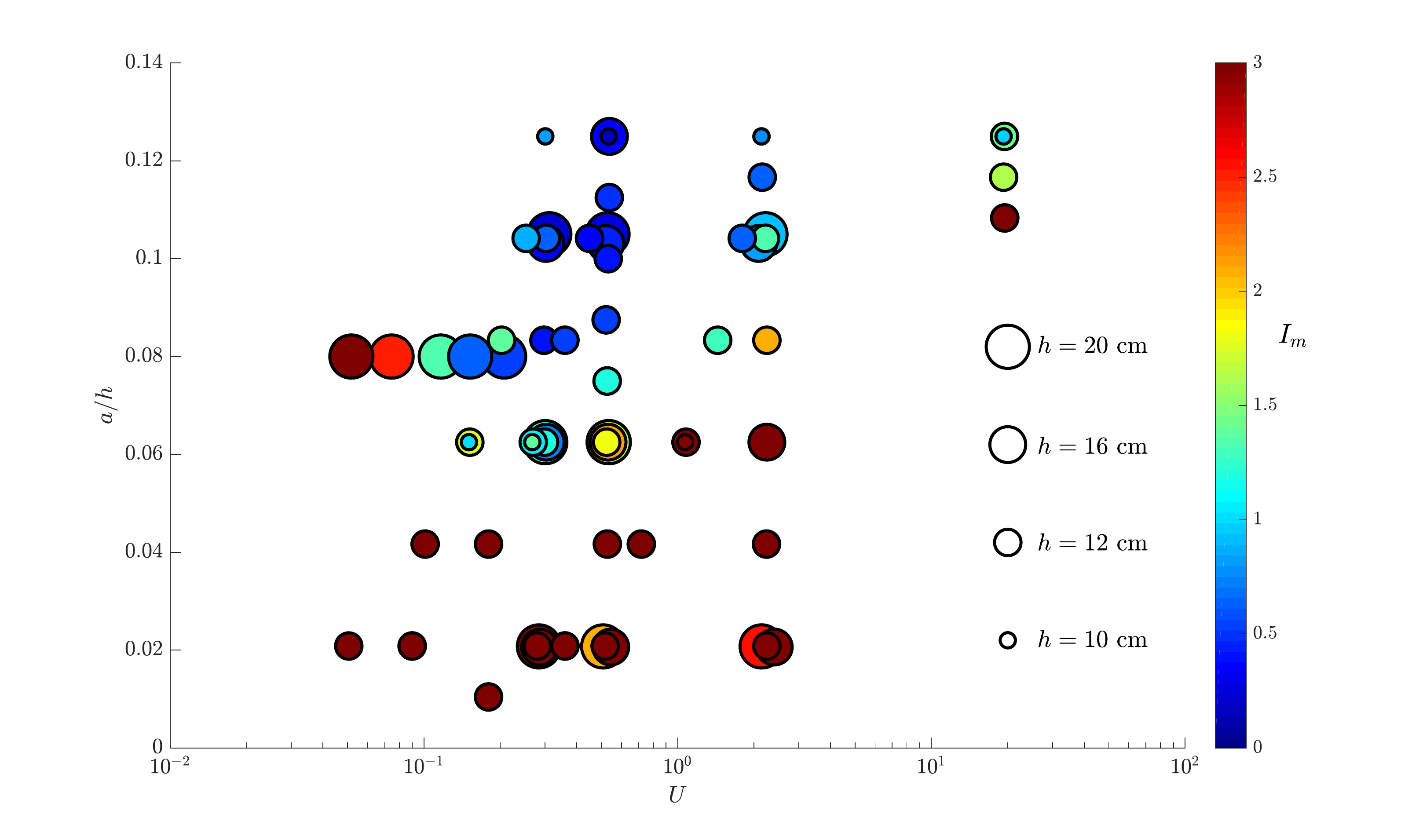}}
\caption{Phase plane of the flume experiments. Forcing wave non-dimensionless parameters: $U$ and $a/h$. The size of the circles relate to the water depth $h$
and the color scale to the value of the randomness index $I_m$ defined by (\ref{index:random}). Values of $I_m < 0.9$ correspond to soliton gaz regimes.}
\label{diag}
\end{figure*}

We conclude from these observations that another way to synthetically discriminate the various regimes, within our experimental framework, is to compute the frequency spectrum as given in Fig.~\ref{PSDU}. In the periodic case, the spectrum is mostly made of narrow peaks with the fundamental peak at the forcing frequency and the other peaks corresponding to higher harmonics. In the random case the spectrum is quite distinct and it is seen to be continuous 
with a flat plateau \cite{pelinovsky2006,RedorPRL} that extends from the forcing frequency down to the smallest resolved frequency. 
At frequencies higher than the forcing frequency the spectrum decays exponentially. 
At this point, a question arises on how to quantify the state of the wave field. At this end, we define a dimensionless randomness index $I_m$ based on the shape of the power spectrum:
\begin{equation}
I_m=\log_{10}\left(\frac{E_{forcing}}{E_{av}}\right) \label{index:random}
\end{equation}
where $E_{forcing}$ is the spectrum value at the forcing frequency (for a given value of the duration of the signal, chosen at 60~s here) and $E_{av}$ is the average value of the spectrum on the range of frequencies lower than that of the forcing. 
For a periodic signal this index is large since the spectrum is strongly peaked at the forcing frequency. In the random case, no peak is present at the forcing frequency and the spectrum at lower frequencies is flat containing most of the energy making the index very small. 

The phase diagram in Fig.~\ref{diag} compiles the various experiments we conducted. The size of the circles indicates what the water depth $h$ is and the color filling the value of the index $I_m$. Blue colors below $1$ can be considered as random states (integrable turbulence) and red/brownish colors are organized states with a certain degree of periodicity. We need to emphasize that we restricted ourselves to relatively low levels of nonlinearity to prevent wave breaking both at the wave generation but also by wave interaction within the flume. 
At the highest values of $a/h$ micro-overturning may still have happened occasionally for very large magnitude of the water elevation due to superimposition of many solitons, but without impacting significantly the global state. In this phase diagram a wedge emerges separating organized states from integrable turbulence. The wedge containing random states is for $a/h$ greater than roughly $0.07$ and $U$ larger than roughly $0.2$. The boundaries of the wedge come with uncertainties. A more accurate identification would require other lengthy measurements since each point in the phase plane corresponds to an experiment duration at least 1 hour-long. Furthermore, note that close to the wedge boundaries, experiments with very close $a/h$ and $U$ values but with distinct $I_m$ (distinct colors) are nearly superimposed suggesting subdominant dependencies to other physical parameters. One obvious extra parameter is the length of the flume (see the discussion in section~\ref{transition}). The points at the far right with $U \simeq 19.3$ that lie in the wedge should lead to fission in soliton trains contradicting an index $I_m$ value above $1$. In these experiments the wavelength of the forcing wave is roughly $10\,$m. Each wave length is too long to completely fission into solitons before the end wall reflection: solitons never really separate before they are dissipated by friction.
At the other end of the phase diagram experiments with $a/h = 0.08$ and $h = 20\,$cm are situations of intermediate water depth with $k \, h > 0.314$ not prone to soliton generation and therefore to the generation of integrable turbulence.

%
\section{Statistics of the stationary regime}
\label{statistics}

In this section statistical distributions and moments of the experimental integrable turbulence are presented. Statistical information complements PST analysis. PST in our present study is a key tool to assess the existence of a soliton gas and characterize the modal content of the gas. Nonetheless integrable turbulence encompasses also shallow water situations where solitonic modes coexist with radiation modes. 
As discussed in the introduction the literature on the statistical description of integrable turbulence is sparse while 
such situation can be present in field measurements \cite{Costa}. A noticeable exception is the numerical study by \cite{pelinovsky2006} on KdV random wave fields, a reference study on these aspects.

\begin{figure}
\centering
\resizebox{0.35\textwidth}{!}{\includegraphics[trim = 0cm 0cm 0cm 0cm, clip]{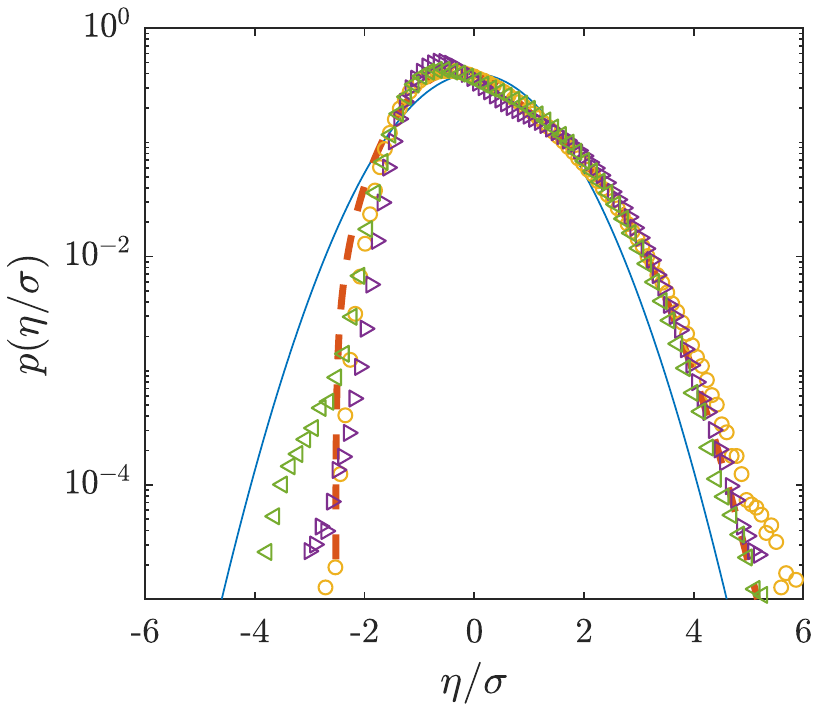}}
\caption{Free surface elevation $\eta$ probability density function computed over $175$ points (every $8\,$cm from $x=9.53$ m up to $x=23.45$ m) and for $25\,$mn in the stationary regime.
The thin blue line is the Gaussian distribution. The thick red dashed line is the  Gaussian distribution corrected with the skewness of the total wave record. Case $U=0.53$, $f=0.6\,$Hz, $a=1.5\,$cm,  standard deviation  $\sigma_t=1.2\,$cm, $\sigma_r=0.98\,$cm, $\sigma_l=0.69\,$cm for resp.
the total time series (\mbox{\large $\circ $}), the right-running part of the time series (\mbox{\large $\rhd$}) and the left-running part of the time series (\mbox{\large $\lhd$}).}
\label{eta:distrib}
\end{figure}

In Fig.~\ref{eta:distrib} we plot the free surface displacement empirical probability density distribution for $U=0.53$ (case of Figs. \ref{spectre} and \ref{separation} and bottom-right-hand subplots in Figs. \ref{PSDU} and \ref{PISTU}). On the same plot the Gaussian 
distribution is the narrow-band linear sea state theoretical distribution \citep{longuethiggins1952}.
Our $U=0.53$ is very close to the Ursell $Ur=0.95$ of \cite{pelinovsky2006} (different definition) and will serve for comparison. In Fig.~\ref{eta:distrib}, the three different probability distributions correspond to the total time series, the right-running and left-running waves. Clearly each of these distributions is positively skewed (higher area under the distribution to the right of the mode). The positive skewness is a well known feature of non-linear waves with peaky crests and wide shallow troughs. The left-running wave elevation distribution tends to be more symmetric as expected due to dissipation. Indeed, the left-running wave are those reflected at the end wall travelling a longer distance from the wavemaker compared to right-running waves. By the Gram-Charlier expansion \citep{longuethiggins1952} including the skewness, the Gaussian distribution is corrected and fits the experimental data fairly well.

\begin{figure}
\centering
\resizebox{0.35\textwidth}{!}{\includegraphics[trim = 0cm 0cm 0cm 0cm, clip]{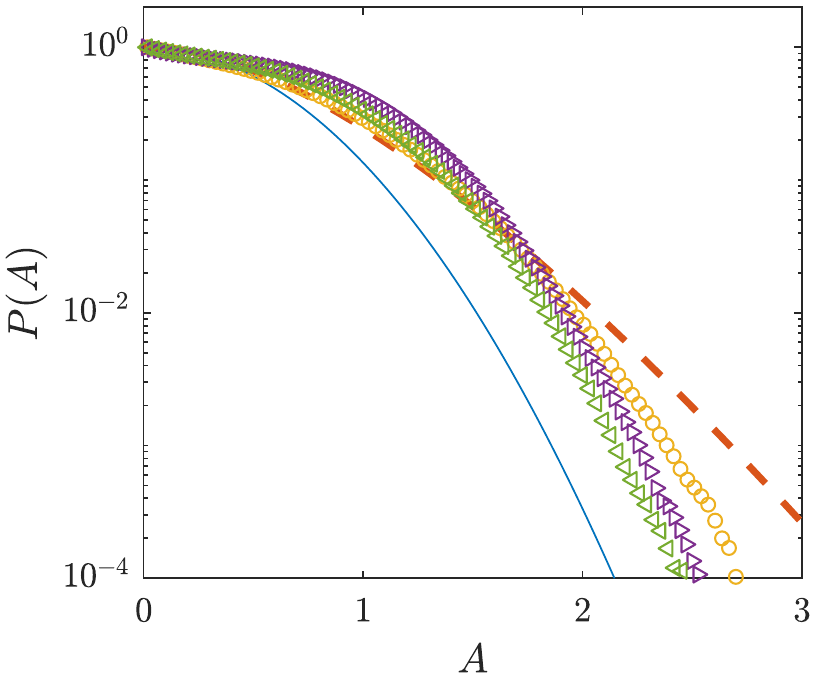}}
\caption{Crest elevation $A$ exceedance probability distribution. Thin blue line: Rayleigh distribution. Thick red dashed line: skewness corrected Rayleigh distribution (\ref{EP:rayleigh}).
Case $U=0.53$, $f=0.6\,$Hz, $a=1.5\,$cm,  standard deviation $\sigma_t=1.2\,$cm, $\sigma_r=0.98\,$cm, $\sigma_l=0.69\,$cm for resp.
the total time series (\mbox{\large $\circ $}), the right-running part of the time series (\mbox{\large $\rhd$}) and the left-running part of the time series (\mbox{\large $\lhd$}).}
\label{crest:exprob}
\end{figure}

For a narrow-band Gaussian linear sea state, the Rayleigh distribution rules the distribution of crest levels \citep{longuethiggins1952}. 
Crest level distributions are also used by \cite{pelinovsky2006} to characterize their random nonlinear shallow water wave field obtained numerically in the KdV framework. The Rayleigh one parameter probability density function $p(A)$ and exceedance probability distribution $P(A)$ write,

\begin{eqnarray}
P(A) & = &  e^{\displaystyle{-2 A^2}} \label{EP:rayleigh} \\
p(A) & = &  4 \, A \, e^{\displaystyle{-2 A^2}} \label{pdf:rayleigh} \\
A & = &  \frac{A_c}{A_s} \\
A_s & = & 2 \, \sigma 
\end{eqnarray}
where $A_c$ are the crests levels determined by a zero-crossing procedure \cite{tucker2001waves}, $\sigma$ the standard deviation of the free surface displacement \cite{pelinovsky2006}. 
Both the experimental and linear wave theoretical exceedance probability are plotted in Fig.~\ref{crest:exprob} for the $U=0.53$ case of soliton gas. As already noticed by \cite{pelinovsky2006} the experimental crest exceedance probability lies above the theoretical Rayleigh probability function. The larger crests tend to be more frequent in an integrable turbulence than in a random field of waves complying to the Rayleigh distribution. Among the larger crests are those of course of the solitons that populate this integrable turbulence.
The present $U=0.53$  close to the $Ur=0.95$ of \cite{pelinovsky2006} shows that the experimental exceedance probability tallies quite nicely with the numerical one (see figure~10 of \cite{pelinovsky2006}). The correction to the Rayleigh distribution suggested by \cite{fedele2009nonlinear} (their equations (5.19) et (5.20)), which takes into account the free surface displacement skewness, is also plotted. While this correction matches the experimental values for small $A$, it overestimates the Rayleigh distribution for large amplitudes corresponding to the large solitons. 
It would suggest that higher order moments are important. 
The moments that characterize best the empirical distribution in Fig.~\ref{eta:distrib} are the skewness ${S}$ and the kurtosis ${K}$, defined for $N$ values equally spaced in time by,
\begin{gather}
{S} = \frac{\mu_3}{\sigma^3} \\
{K} = \frac{\mu_4}{\sigma^{4}} \\
\sigma^2 = \frac{1}{N} \, \sum_i \left( \eta_i - \overline{\eta} \right)^2;  \\
\mu_3 = \frac{1}{N} \, \sum_i \left( \eta_i - \overline{\eta} \right)^3; \ \ \ \
\mu_4 = \frac{1}{N} \, \sum_i \left( \eta_i - \overline{\eta} \right)^4 
\end{gather}
Where $\overline{(.)}$ stands for the time average. For a Gaussian random wave field ${S} = 0$ and ${K} = 3$. The kurtosis measures the heaviness of the probability distribution tails while the skewness indicates the asymmetry of the distribution around the average. The experimental values for the skewness and the kurtosis for the left-running waves in the $U=0.53$ case are ${S}= 0.70$ and ${K}= 3.40$. The corresponding values of \cite{pelinovsky2006} are ${S}= 0.73$ and ${K}= 3.45$. Our experiments and their numerical simulations yield very close values. We compared their numerics with the experimental left-running wave field because the latter is deemed to have achieved statistical stationarity. While our experimental procedure starting from a sine wave differs from the initial conditions of \cite{pelinovsky2006}, both generate a very similar long term integrable turbulence stationary state
which is mainly characterized by the Ursell number.

\begin{figure}
\centering
\resizebox{0.35\textwidth}{!}{\includegraphics[trim = 0.6cm 4.2cm 0cm 4.5cm, clip]{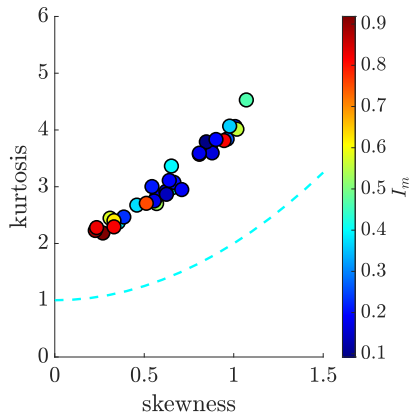}}
\caption{Skewness ${S}$ of the free surface elevation versus the kurtosis ${K}$. Colors refer to the level of the mixing index $I_m$. The dashed line is the lower bound of the Kurtosis: ${K} \ge 1 + {S}^2$ (\citep{pearson1916} p.432). Data corresponding to points in Fig.~\ref{diag} with $I_m <0.9$.}
\label{fig:skewkurt}
\end{figure}

In their numerical study, \cite{pelinovsky2006} show that for $U> 0.16$, both the skewness and the kurtosis increase quasi-linearly with the Ursell number (see their figure~7) suggesting that the skewness and the kurtosis are linearly related. We show in Fig.~\ref{fig:skewkurt} that this also stands in our experiments. In this figure all the experimental runs labeled as integrable turbulence, that is with a mixing index $I_m < 1$, are plotted and clearly align.

\begin{figure}
\centering
\resizebox{0.35\textwidth}{!}{\includegraphics[trim = 0cm 0cm 0cm 0cm, clip]{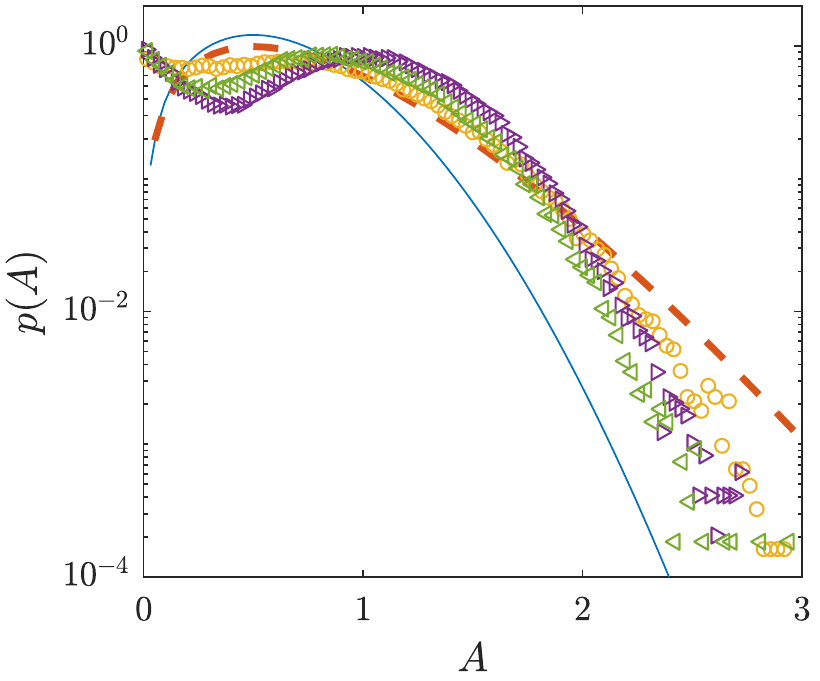}}
\caption{Crest elevation $A$ probability density function. Thin blue line: Rayleigh pdf (\ref{pdf:rayleigh}). Thick red dashed line: skewness corrected Rayleigh pdf \cite{fedele2009nonlinear}.
Case $U=0.53$, $f=0.6\,$Hz, $a=1.5\,$cm,  standard deviation $\sigma_t=1.2\,$cm, $\sigma_r=0.98\,$cm, $\sigma_l=0.69\,$cm for resp.
the total time series (\mbox{\large $\circ $}), the right-running part of the time series (\mbox{\large $\rhd$}) and the left-running part of the time series (\mbox{\large $\lhd$}).}
\label{crest:pdf}
\end{figure}

\begin{figure*}
\centering
\resizebox{0.78\textwidth}{!}{\includegraphics[trim = 2.9cm 1.7cm 3.3cm 1.6cm, clip]{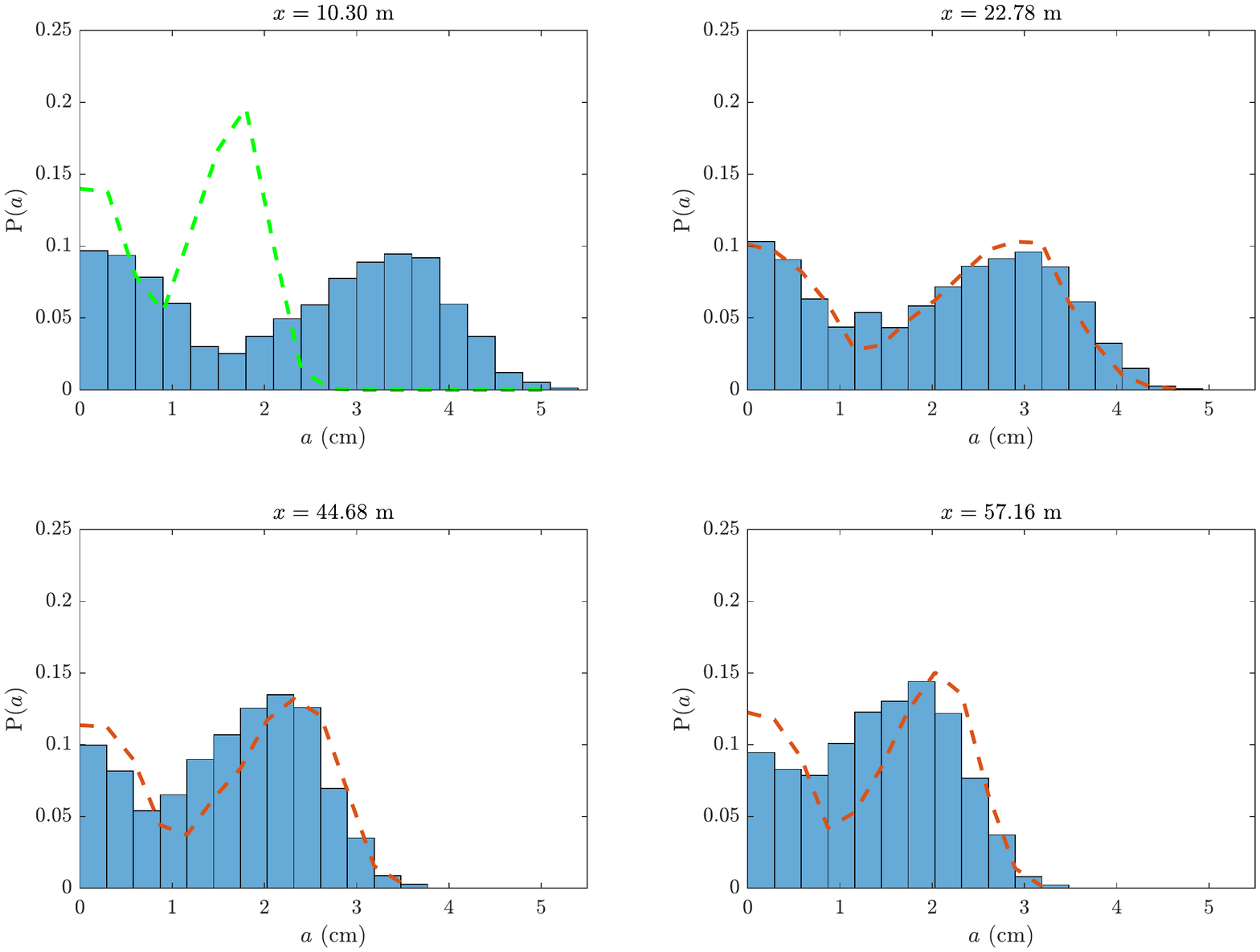}}
\caption{PST soliton amplitudes histograms at different locations in the flume. Top panels correspond to right-running waves. Bottom panels correspond to left-running waves. Histograms with bars: PST processed experimental data; dashed lines predicted histograms by applying the Keulegan damping law (\ref{Keu48}); green dashed: propagated histogram from $x=57.16\,$m to $x=10.3\,$m as if the wavemaker was a fixed vertical boundary. Case $U=0.53$, $f=0.6\,$Hz, $a=1.5\,$cm and $\nu = 1.2 \, 10^{-6} \, {\rm m^2/s}$}
\label{sol_ampli:hist}
\end{figure*}

The pdf of the crests levels is plotted in Fig.~\ref{crest:pdf}. At small amplitude this pdf shows that the empirical distribution exhibits a gap instead of a maximum around 
the crest value of $0.5$. The data departs from the Rayleigh pdf and this discrepancy is badly explained with the skewness modified Rayleigh pdf.
It appears that crests levels follow a kind of bimodal distribution. Some of the crests detected are probably those of solitons but not all of them.
Consequently the latter statistical characterizations do not inform on how close a given state of integrable turbulence is to a soliton gas. 
In contrast the PST provides quantitative information on the solitonic content of the times series. Of interest are the statistics of the amplitudes of the solitonic modes that relate to the finite-band spectrum of the Schr\"odinger equation in the PST. In Fig.~\ref{sol_ampli:hist} the  empirical amplitude probability histograms are plotted at $4$ different locations along the flume for the Ursell number $U=0.53$ case. The PST is run on $55$ overlapping time segments of $60\,$s for right-running waves at $x=10.3\,$m and $x=22.78\,$m and for left-running ones at $x=44.68\,$m and $x=57.16\,$m. The time window of $60\,$s is slightly smaller than the time necessary for a wave to travel twice a total flume length ($62.2\,$s for a depth of $12\,$cm ) which includes a reflection on the end wall and one on the wavemaker. This ensures we process a time series excluding waves measured twice. 

The empirical histograms all exhibit $2$ wide peaks of solitonic modes as if the distributions were the combination of two distributions that characterize the experimental integrable turbulence,
an aspect pointed out in Fig.~\ref{crest:pdf}. 
As the wave trains travel back and forth, both the width of the histograms and the largest amplitude peak of the distribution decrease. In the ideal KdV integrable turbulence
these distributions should be invariant with $x$.
In the experiments as the solitons propagate, dissipation progressively reduces their amplitude according to the Keulegan dissipation ``law''  \citep{keulegan1948}. The Keulegan dissipation law reads,
\begin{gather}
 \left(\frac{a}{h}\right)^{-1/4} = \left(\frac{a_0}{h} \right)^{-1/4} + \, D \, \frac{x}{h}  \label{Keu48} \\
 D = \frac{1}{12} \,\left(1+ \frac{2 \, h}{w} \right)\, \sqrt{ \frac{\nu}{g^{1/2} \, h^{3/2}}}
\end{gather}
where $w$ is the width of the flume, $\nu$ the water viscosity and $a_0$ the initial soliton amplitude.
This Keulegan law has been thoroughly validated experimentally \cite{keulegan1948,renouard1985,RedorEIF}.
In the dissipation process the large solitons become intermediate amplitude solitons that tend to populate the initial gap around $1.5\,$cm. Moreover as the largest decrease more rapidly than the smallest, as given by the Keulegan dissipation ``law'' (\ref{Keu48}), the distribution is severely eroded from the right end of large amplitudes. In Fig.~\ref{sol_ampli:hist} the initial distribution of right-running waves at $x=10.3\,$m is transformed into the distribution at the next location ($x=22.78\,$m) by applying Keulegan amplitude reduction law (\ref{Keu48}). The initial distribution is likewise  ``transformed'' to the distributions at other locations up to and back to $x=10.3\,$m. It is assumed that only reflections take place at the end wall and at the wavemaker.
The successive transformed distributions compare well with the measured ones except at $x=10.3\,$m. At this location the measured wave distribution is much wider with peak values at higher amplitudes than that of the Keulegan-transformed distribution. This indicates that the wave train gains energy as it bounces back on the wavemaker. On one hand it is due to interactions of the various waves incoming on the moving wavemaker and on the other hand to the generation of new waves by the wavemaker.  By positive or negative feedback some waves have an amplitude that increases and the contrary for others. The net result is an increase in the wave energy flux at $x=10.3\,$m. 
This is confirmed by PST analyzing this  $U=0.53$ sinusoidal wavemaker forcing wave. The PST yields one set of $n$ modes of modulus larger than $0.99$, thus in practice a set of $n$ amplitudes very close to $2.84\,$cm. Nonlinear modes of lower modulus are also detected in this PST processing. Thus, the modal content of the forcing sinusoid is one solitonic mode and at least one cnoidal type mode per period. The solitonic mode amplitude is smaller than that of the highest amplitude peak at $x=10.3\,$m which is around $3.5\,$cm. The likely explanation is that in the stationary regime the nonlinear modes of the forcing wave interact positively with the moving wavemaker to a point where the solitonic mode amplitude increases and the most nonlinear cnoidal mode becomes a solitonic mode of modulus $m>0.99$. The feedback of the wavemaker on the travelling waves produces $2$ sets of solitonic modes.

The mechanism by which soliton content and amplitudes change, are now discussed in more detail by analyzing the transient route to integrable turbulence.
%
\section{Transition to integrable turbulence: randomization of the periodic forcing}
\label{transition}

\begin{figure*}
\centering
\resizebox{0.8\textwidth}{!}{\includegraphics[trim = 2.5cm 1.6cm 3.3cm 2.1cm, clip]{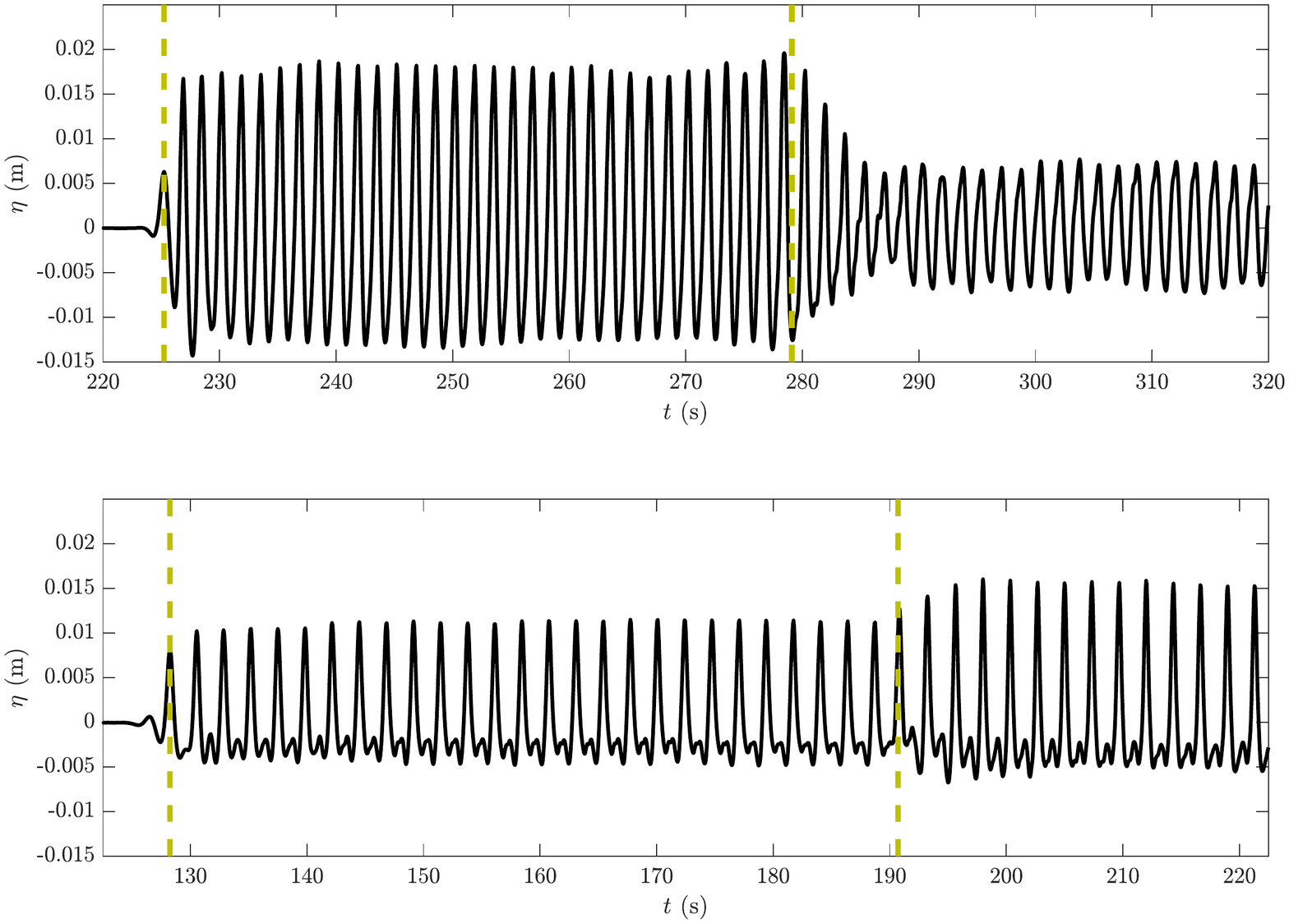}}
\caption{Attenuation and amplification of right-running wave trains. Initial free surface elevation $\eta$ at $x=10.5$~m. Top panel: $a=1.5$~cm, $f=0.6$~Hz ($T=1.67$~s), water depth $h=16\,$cm, $a/h=0.094$ and $U=0.30$.
bottom panel: $a=0.75$~cm, $f=0.43$~Hz ($T=2.33$~s),  water depth $h=12\,$cm, $a/h=0.062$ and $U=0.53$.  Vertical dashed lines separated by the time necessary for a wave to acheive a round trip of $2 \times 33.73\,{\rm m}$ at $c_0$.}
\label{first1}
\end{figure*}

In a periodic box and with an initial condition made of a sine wave, Zabuski \& Kruskal \citep{zabusky1965interaction} numerically predict a recurrence, i.e. the fact that the wave system retrieves in a finite time a state very close to the initial condition \cite{zabusky1965interaction}. In our experimental setup, it is not possible to start from an arbitrary initial condition which is not rest. What can only be achieved in a controlled way, is to start wave forcing at one end with a wavemaker and a body of water at rest in the entire flume. 

The reflection on the moving wavemaker is a key point to the transition towards integrable turbulence. Indeed, a wave that travels from and back to the wavemaker, after reflection on the end fixed wall, will interfere with the wavemaker motion. Depending on the relative phase of the wave with respect to that of the wavemaker, the wave amplitude may increase or decrease. A  well known situation is that of a standing wave. In Fig.~\ref{first1} two experimental examples of such positive or negative interference are given. The wave trains plotted in Fig.~\ref{first1} are right-running filtered waves. To assess if an amplification or attenuation takes place the one way filtering is necessary to suppress any partial standing wave pattern, with partial nodes and anti-nodes, which blur the right running wave amplitudes. The bottom panel is that of an amplification for $h=12\,$cm ($c_0 = 1.08\,{\rm m/s}$). The wave crest marked with a dashed vertical line at $t=128.25$~s travels away and a complete round trip to be recorded as a right running wave at $x=10.5$~m at $t=190.7$~s. This crest is almost in phase with a newly generated wave producing a positive interference.

By contrast in the top panel, the right-running waves undergo an attenuation for $h=16\,$cm ($c_0 = 1.25\,{\rm m/s}$). The wave crest marked with a dashed vertical line at $t=225.25$~s travels away and a complete round trip to be recorded as a right running wave at $x=10.5$~m at $t=279.1$~s. The wave crest after a complete round trip is out of phase with a newly created wave (in phase with the trough) producing a negative interference
and therefore an attenuation. In section \ref{parametric} the length of the flume was characterized as a subdominant parameter. Indeed, a slight change of length or equivalently of wavemaker frequency can possibly shift the feedback of the wavemaker with the traveling waves from positive to negative and conversely.

\begin{figure*}
\centering
\resizebox{\textwidth}{!}{\includegraphics[trim = 0.8cm 2.3cm 0.5cm 0.6cm, clip]{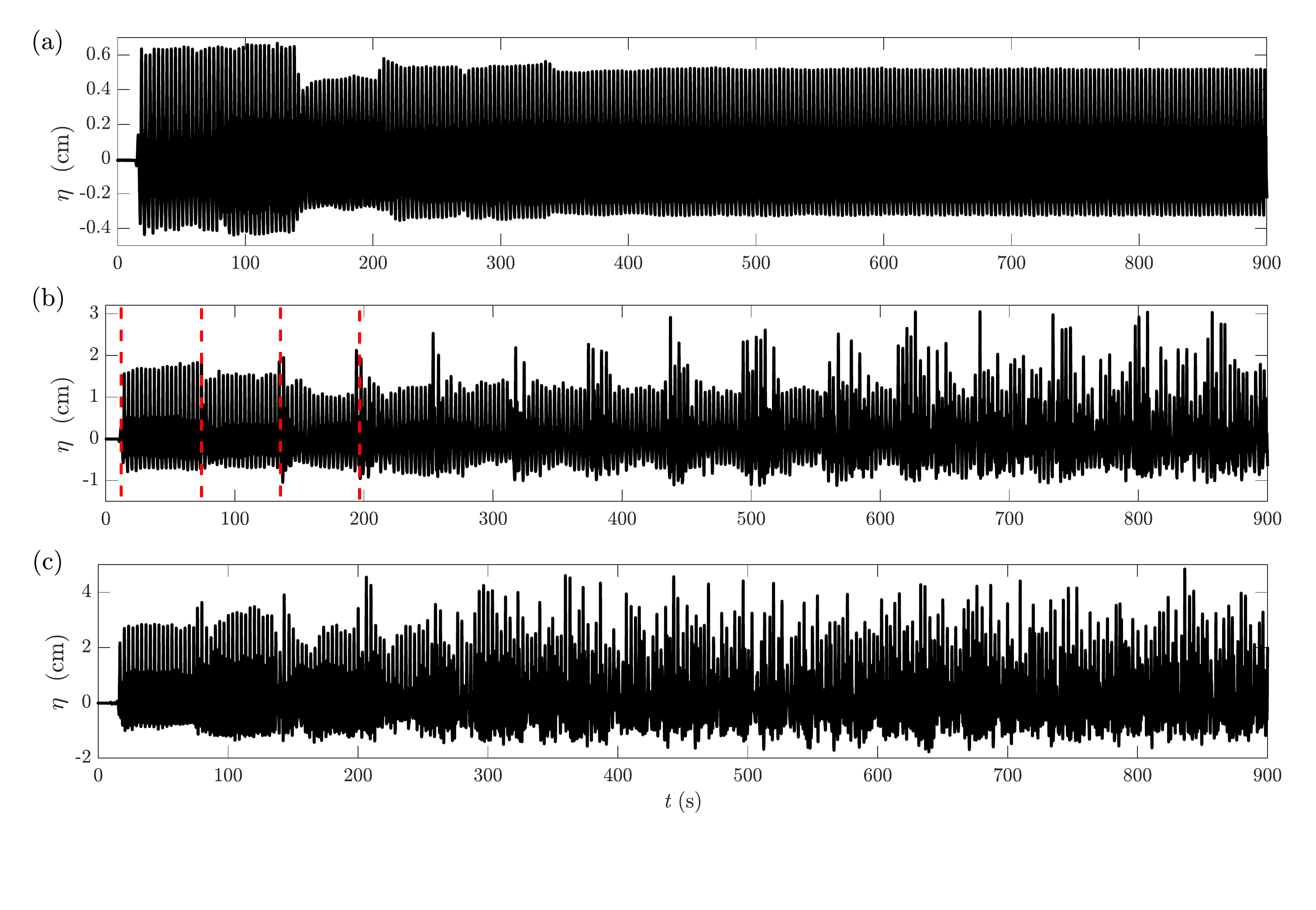}}
\caption{Transition to integrable turbulence. Free surface elevations $\eta$ at $x=10.5$~m for 3 different wave amplitude forcing. Wave-maker frequency $f=0.3$~Hz and water depth at rest $h=12$~cm. Top panel (a) $a= 0.5$~cm, $a/h=0.04$ and $U=0.72$; middle panel (b):  $a= 1.0$~cm, $a/h=0.08$ and $U=1.44$, vertical dashed lines separated by the time necessary for a wave to complete a round trip in the flume; bottom panel (c):  $a= 1.5$~cm, $a/h=0.12$ and $U=2.16$.}
\label{first}
\end{figure*}

Fig.~\ref{first} shows water elevation records at $10.5$~m from the wavemaker with three conditions of forcing with same frequency and water depth but distinct amplitudes and thus distinct values of nonlinearity level $a/h$ and Ursell number $U$. 

In the first case (a), for the lowest value of $a/h$, wave crest elevations evolve by discrete set of steps (either increasing or decreasing) every round trip (every $62.2$~s), but on the long run the signal becomes periodic and the crest levels are smaller than initially. The soliton content processed by PST gives $1$ soliton of amplitude $a_s = 0.76\,$cm 
per period in the initial first round trip (either with threshold $m=0.99$ or $m=0.999$) and no solitons at all in the periodic tail of the signal. The negative feedback produced by the wavemaker inhibits the survival of solitons.

In the intermediate case (b) even though at each round trip the waves overall are slightly attenuating for $t<200\,$s there are small time intervals of the signal that show irregular crest level increase. These time intervals occur roughly every round trip of $62.2\,$s, indicated by the  vertical dashed red lines and widen at each round trip. These amplification sequences eventually overlap after $t = 700\,$s to produce a random state. 
The PST analysis, at the threshold $m =0.99$, of $6$ periods in the first round trip gives $12$ solitons that is $2$ per period with average amplitudes of respectively $a_{s1} = 2.25\,$cm and $a_{s2} = 1.76\,$cm. The equivalent analysis on sequences of $6 \times T$  duration beyond $t = 700\,$s yields an average of $16$ solitons with the largest amplitude at $a_{s} = 3.9\,$cm. 

In case (c), the disorganization is even faster than in case (b). The PST analysis, at the threshold $m=0.99$, of $6$ periods in the  first round trip gives $3$ solitons per period with amplitudes of respectively $a_{s1} = 3.95\,$cm $a_{s2} = 2.43 \,$cm and $a_{s3} = 0.92 \,$cm. The equivalent analysis on sequences of $6 \times T$  duration beyond $t = 2750 \,$s yields an average of $21$ solitons with the largest amplitude at $a_{s} = 6.1\,$cm. 

The former last two cases clearly indicate on the one hand that even though the initial wave trains contain solitons the stationnary wave state contains more, with larger amplitudes, a signature of the wavemaker importance in the generation by sinusoidal forcing of integrable turbulence. Indeed, would the wavemaker make no difference, integrability would impose the same number and amplitudes of solitons at the beginning and in the stationary wave state. On the other hand, they emphasize the possible role of the first soliton train. Indeed, the soliton amplitudes of this first train are smaller than the subsequent ones, but they travel on the rest level (same as the mean level $h$ in our experiments). As it appears on Fig.~\ref{fission1} the other soliton trains propagate on a lower reference water level and with adverse velocity \citep{osborne1986solitons,Redorthesis}. The difference $\Delta h$ between mean water level and the reference water level roughly corresponds to the mass of the solitons contained in one forcing wave wave length divided by the wave length $l$. 
As \cite{osborne1986solitons} showed the velocity $c$ of a soliton propagating on a reference level different from the rest level is,
\begin{equation}
    c = \sqrt{g \, h} \, \left(1 + \frac{1}{2} \frac{a}{h} - \frac{3}{2} \frac{\Delta h}{h}  \right).
\end{equation}            
Consequently the very first train travels faster, it is not synchronized with the others and can be amplified by the wavemaker. In this case when large enough, the first soliton induces large phase shifts that in turn can disynchronize other solitons that can be either amplified or attenuated by the wavemaker. This chain of interactions triggers the randomization of the wave motion by shuffling solitons that initially were ordered by decreasing amplitudes. Since phase shifts between soliton increase with amplitude and thus with the nonlinear $a/h$ parameter it conceivably explains why case (c) disorganizes more rapidly than (b). The first soliton train acts as a catalyst for the integrable turbulence.

\section{Conclusion}

Our experimental set-up differs from the integrable framework of Zabusky \& Kruskal \cite{zabusky1965interaction} on various points. The most obvious is dissipation that imposes some continuous energy flux input for the wave motion to possibly reach a statistically stationary wave regime. However, the time scales of dissipation, well represented by the Keulegan law, are much larger than those involving soliton interactions which suggest that integrability still holds locally. Indeed, we show that once a soliton gas is formed dissipation slowly alters the amplitude distribution but not to the point where it would be obliterated in a flume length propagation time. 

The second difference lies of course in the finite flume length and the reflecting behavior of both ends of the flume. We show that it does not only allow for bi-directionnal integrable turbulence as described by the Kaup-Boussinesq equations \cite{kaup1975higher,nabelek2020} but also produces desynchronization of the waves with the wavemaker motion. 

Finally the third difference and probably the most important for the generation of integrable turbulence is the wavemaker feedback on wave amplification and attenuation. We emphasize that the route to a random integrable turbulence wave motion strongly depends on the degree of non-linearity $a/h$ of the wavemaker motion since the energy flux input by the wavemaker is proportional to $(a/h)^2$. 

To characterize if a given wave motion state is close to integrable turbulence or not, we suggest different metrics.
The nondimensionnal wavemaker Ursell number and nonlinear parameter $a/h$ are relevant numbers to delineate integrable turbulence from other regimes. An experimental coverage of the ($U$, $a/h$) plane indicates that integrable turbulence in our settings will be sustained if $a/h > 0.07$ and $U > 0.2$. These bounds are approximate and should be confirmed with more experiments.
The mixing index $I_m$ that characterizes the shape of the power density spectrum is used to assess if wave periodicity has disappeared in the final wave motion state, one of the signatures of integrable turbulence.   
We sustain that, alike numerical simulation based on the KdV equation, kurtosis increases quasi linearly with skewness. This needs to be further investigated in other conditions, but such property may be useful in the analysis of oceanic wave data.
Finally the periodic scattering transform (PST) is used to precisely characterize the spectral content of the empirical integrable turbulence. This transform is unavoidable to determine the solitonic and radiative  content of a given regime and thus is the only tool to supply soliton amplitude distributions and fully characterize wave motion random states.

The difference between unidirectional and bi-directionnal integrable turbulence has not been addressed in the present study. 
The Kaup-Boussinesq equations \cite{kaup1975higher} offer an integrable framework to describe bi-directionnal integrable turbulence in which the interactions between counter-propagating solitons 
generate pulses than the sum of the amplitude of the two solitons before interaction \cite{seabra1989weak,chen2014laboratory} and also phase shifts. Nonetheless the approach used in the present study takes advantage in the fact that, for weakly non-linear solitons in an integrable turbulence, the inverse scattering problem for the Kaup-Boussinesq equations can asymptotically be decomposed in left-going and right-going  KdV inverse scattering problems \cite{kaup1975higher}. There are some indications in the literature of IST techniques for Kaup-Boussinesq \cite{kamchatnov2003asymptotic}. However, a practical periodic IST is yet to be developed and would require extensive numerical developments. In future work different experimental boundary conditions could be considered such as absorbing conditions at the end wall with re-injection of the outgoing waves at the wavemaker simultaneously with the generation of new wave trains conducive to a one way integrable turbulence. While easy to implement numerically we foresee some technical difficulties to do it on a real-time basis.

%
\begin{acknowledgments}
We thank {S. Randoux} and {P. Suret} (PhLAM, Universit\'e de Lille) for the fruitful discussions during I. Redor's PhD.
This project has received funding from the European Research Council (ERC) under the
European Union’s Horizon 2020 research and innovation program (Grant No. 647018-WATU). It was also partially supported by the Simons Foundation through the Simons collaboration on wave turbulence. 
\end{acknowledgments}
\providecommand{\noopsort}[1]{}\providecommand{\singleletter}[1]{#1}%
%

\end{document}